\documentclass[aps,11pt,preprint,superscriptaddress,showkeys,nofootinbib]{revtex4}

\usepackage[dvips]{graphicx}
\usepackage{amsmath}
\usepackage{amsfonts}
\usepackage{amssymb}
\usepackage{mathrsfs}
\usepackage{bm}
\usepackage{verbatim}

\newcommand{\beq}{\begin{equation}}
\newcommand{\eeq}{\end{equation}}
\newcommand{\bea}{\begin{eqnarray}}
\newcommand{\eea}{\end{eqnarray}}

\newcommand{\refeq}[1]{Eq.~\eqref{#1}}
\newcommand{\reffig}[1]{FIG.~\ref{#1}}

\newcommand{\tr}[1]{\text{#1}}

\newcommand{\iv}[1]{\bm{#1}}

\newcommand{\mqcd}{M_{\textrm{QCD}}}

\graphicspath{{../figs/}{../plots/}}

\begin{document}

\title{$\pi N$ Scattering in the $\Delta(1232)$ Region
in an Effective Field Theory}

\author{Bingwei Long}
\email[Corresponding author. Tel: +39(0461)314-776; Fax:
  +39(0461)935-007; Email: ]{long@ect.it}
\affiliation{European Centre for Theoretical Studies in Nuclear
  Physics and Related Areas (ECT*), Strada delle Tabarelle 286,
  I-38100 Villazzano (TN), Italy}
\affiliation{Department of Physics, University of Arizona, 1118 East 4th
  Street, Tucson, Arizona 85721, USA}
\author{U. van Kolck}
\email[Email: ]{vankolck@physics.arizona.edu}
\affiliation{Department of Physics, University of Arizona, 1118 East 4th
  Street, Tucson, Arizona 85721, USA}

\date{\today}

\begin{abstract}
We develop a generalized version of
heavy-baryon chiral perturbation theory to describe 
pion-nucleon scattering in a kinematic
domain that 
extends continuously from threshold to the delta-isobar peak. 
The $P$-wave phase shifts are used to illustrate this framework.
We also compare our approach with those in the literature
that concern the delta resonance.
\end{abstract}



\preprint{ECT*-09-08}
\preprint{INT-PUB-09-036}

\maketitle

\section{Introduction\label{sec:intro}}

Pion exchange provides the long-range components of nuclear forces,
and crucial to its understanding is pion-nucleon ($\pi N$) scattering. 
A prominent feature of the latter is 
the delta resonance, $\Delta(1232)$, a peak in the elastic
cross section at the center-of-mass (CM) energy 
$m_{\Delta} \equiv m_N + \delta \simeq 1230$ MeV,
where $\delta \sim 290$ MeV is the nucleon-delta mass splitting \cite{pdg}. 
Our goal here is to
investigate $\pi N$ scattering from threshold up to the delta resonance
in an effective field theory (EFT).

It is well known that resonances can be 
studied by considering the unitarity and analyticity of the
$S$ matrix. Assuming for simplicity that
there is only one decay channel for a resonance, when the
CM energy $E$ is close enough to the resonance, the $T$-matrix element in the
resonance channel can be 
written in the form of a Breit-Wigner formula plus a non-resonant
background (see, \textit{e.g.},
Refs.~\cite{colltheo,weinbergbook}),
\beq
T(E) \equiv -i\left\{\exp\left[2i\theta(E)\right] - 1\right\} =
  -\frac{\Gamma}{E - E_R + i\Gamma/2} \left(1 + i T_B\right) + T_B \;
, \label{eqn:bwback}
\eeq
where $\theta(E)$ is the phase shift, $E_R$ and $\Gamma/2$ are 
real numbers that represent the energy and half-width of the resonance
(that is, the pole position in the complex-energy plane), 
and $T_B$ is a complex number
often called the non-resonant background amplitude,
which can be written as $T_B=-i[\exp(2i\theta_0) - 1]$ in terms
of a non-resonant phase shift $\theta_0$.

Though based on general principles, the direct application
of \refeq{eqn:bwback} relies on a few assumptions. First, 
in many cases where $E_R$ and $\Gamma$ are unknown \textit{a priori},
they are free parameters of \refeq{eqn:bwback}
despite the fact that they are related in some underlying theory (which
is Quantum Chromodynamics (QCD) in the delta
case). In the absence of $T_B$, one could precisely determine $E_R$
by where the phase shift passes through $\pi/2$ (hence a sharp peak in
the cross section). However, $T_B$ is in general non-vanishing so one
has to make an educated guess about where the resonance window is
centered.
Therefore, the values of $E_R$, $\Gamma$, and $T_B$ based on
\refeq{eqn:bwback} may be model-dependent.
Second, one has to assume how close to the resonance is 
``close enough''. 
Particularly, is $E_R$ small enough so that \refeq{eqn:bwback}
is valid even at threshold? 
If \refeq{eqn:bwback} works only in the resonance region, one has
to assume that the data points one wants to fit are sufficiently close to the 
resonance. 
Or can we extend \refeq{eqn:bwback} simply by allowing 
$T_B$ and/or $\Gamma$ to be functions of energy (as
it happens in general in field theory) and, if so, 
with what constraints on $T_B(E)$ and $\Gamma(E)$? 
If $\Gamma$ depends on the energy, then
the position of the pole is {\it not} given by $E_R-i\Gamma(E_R)/2$, 
and the physical interpretations of $E_R$ and $\Gamma(E_R)$ are unclear.
In the case of the delta, one can show explicitly \cite{gegelia} that
$E_R$ and $\Gamma(E_R)$ are indeed unphysical
in the sense that they depend at two-loop level on the choice of fields.

Given that one cannot, at present,
straightforwardly solve QCD at 
low energies due to the difficulties posed by
the large strong-coupling constant and the small pion mass $m_\pi$, 
EFT is a good alternative for
describing low-energy nuclear and hadronic physics consistently. Following
several seminal papers \cite{weinberg79},
EFTs have been developed as model-independent
approximations to low-energy strong interactions, which can be
systematically improved by a series in powers of $Q/\mqcd$, where $Q$
refers generically to small external momenta and $\mqcd \sim 1$ GeV
is the characteristic QCD scale.
The loyalty of EFTs to QCD, the
underlying theory of low-energy strong interactions, is manifested by
the fact that EFTs inherit all the symmetries of QCD, among which
chiral symmetry is probably the most nontrivial.
For reviews, see, for example, Refs.~\cite{BerKaiMei,bira99review}.

A particular EFT, chiral perturbation theory (ChPT),
specializes, and has proven quite successful, in
processes involving at most one nucleon \cite{BerKaiMei}.
At energies $E$ close to the $\pi N$ threshold, $E\sim m_\pi$,
that is, for $Q$ 
around the pion mass,
the delta dynamics can be considered short-range physics, 
which amounts to treating $\delta$ as a large scale,
$Q \sim m_\pi \ll \delta$. ChPT with only pion and nucleon fields 
has been extensively applied to $\pi N$ scattering
near threshold \cite{originalS, mojzis, fettes-deltaless-is,otherdeltaless}.
However, it is a perturbative expansion in powers of $Q/\delta$ and 
$m_\pi/\delta$, which should converge slowly because in the real world
$\delta \simeq 2m_\pi$. 
The slow convergence then contaminates pion-exchange nuclear forces 
\cite{vijay},
where pion energies $\omega$ are small compared to the $\pi N$ threshold
energy, $\omega\ll m_\pi$.

One can improve convergence by considering the
delta as an explicit degree of freedom, in which case one can take 
$Q \sim m_\pi \sim \delta$. One can show 
that hadronic scattering amplitudes can
be written as perturbative series in powers of $Q/\mqcd$, 
$m_\pi/\mqcd$, and $\delta/\mqcd$ \cite{jenkins, hemmertdelta},
and that pion-exchange nuclear forces display a good convergence pattern
\cite{bira-thesis,kaiserbrock,vijay}.
The positive role of an explicit delta field in $\pi N$ scattering within ChPT 
has been explored \cite{threshold-delta1} and demonstrated
in a fully consistent calculation \cite{threshold-delta2}.

Nevertheless, this perturbative expansion diverges around the resonance,
where the delta goes on-shell. This is not surprising since
the perturbative nature of 
standard ChPT makes it impossible to
describe such a non-perturbative phenomenon. 
In order to fully describe low-energy $\pi N$ scattering 
one needs to resum certain terms in the expansion so as to have finite results
throughout the low-energy region. 
A non-perturbative treatment of the delta within ChPT was considered in 
Ref.~\cite{ellis-tang}, where the leading delta self-energy
was resummed. 
However, a systematic resummation did not exist until
the seminal work of Ref.~\cite{pascalutsa},
where it was justified 
by a power counting 
based on three separate scales 
$m_\pi \ll \delta \ll \mqcd$.
This idea has since been applied to various electromagnetic
reactions in the 
delta region \cite{pascalutsa,morepascalutsa1,morepascalutsa2},
but for $\pi N$ scattering few results have been published 
\cite{morepascalutsa2}. 

Note that other approaches exist to incorporate
the delta (and other resonances) consistently with chiral symmetry
and field-definition independence.
They are reminiscent of the original approach  \cite{original,bira-thesis}
to nucleon-nucleon interactions using chiral Lagrangians:
a pion-nucleon 
``kernel'' is first derived in ChPT to a certain order and then unitarized,
for example using the $N/D$ method \cite{unitarization1}
or the Bethe-Salpeter equation \cite{unitarization2}.
Power counting is not manifest 
at the amplitude level, but
good results for pion-nucleon phase shifts are obtained past 
the delta region.

In this paper 
we realize the delta as a heavy baryon that fulfills
Lorentz invariance order by order in powers of $Q/m_N$, using the
nonrelativistic delta field as one of the building blocks.
We employ a power counting developed for generic 
narrow resonances \cite{resonances},
previously applied to the shallowest $P$-wave 
resonance in nucleon-alpha particle scattering at very low energies
\cite{resonances, halos}.
In this power counting we consider only
two scales $M_{\tr{lo}} \sim \delta \sim m_\pi$
and $M_{\tr{hi}} \sim \mqcd$, and expand in
$Q/M_{\tr{hi}}$ and $M_{\tr{lo}}/M_{\tr{hi}}$.
Certain contributions are enhanced near the resonance over
the standard ChPT counting, which continues to apply near threshold.
We can calculate the $\pi N$ amplitude spanning the two regions
in a controlled expansion. 
We illustrate the method by explicitly calculating the first three
orders of the expansion and comparing with 
known $P$-wave phase shifts 
\cite{gwpwa} and scattering volumes \cite{matsinos}.
In the delta region our approach is similar, but not identical,
to that in Ref.~\cite{pascalutsa}.
We compare our approach with those in Refs. \cite{ellis-tang,pascalutsa}
as the differences arise.

The rest of the paper is organized as follows.
In Sec.~\ref{sec:el} we
describe the heavy-baryon chiral Lagrangian that has an explicit delta field.
(Some of the details of our implementation of the delta field are
relegated to the Appendix.)
We discuss the power counting in Sec.~\ref{sec:powct}. In
Sec.~\ref{sec:pre} we 
show how several key ingredients of the $\pi N$ amplitude are calculated. 
The calculation and renormalization of the EFT amplitude are carried out 
in Sec.~\ref{sec:amps}. We then fit the $P$-wave phase shifts in
Sec.~\ref{sec:psa}. A summary of our results and a
conclusion are offered in Sec.~\ref{sec:summary}.

\section{Effective Lagrangian\label{sec:el}}

In this section we briefly review how the effective 
Lagrangian is constructed.
Much of this has already been discussed
in the literature (see, \textit{e.g.}, 
Refs.~\cite{weinbergbook, BerKaiMei, bira99review, bira-thesis}),
our review here serving mainly to establish the notation.
Some details about the delta can be found in the Appendix.

\subsection{Fields and symmetries}

The effective Lagrangian, with hadronic degrees
of freedom, is expected to exhibit the symmetries of QCD, which include 
Lorentz invariance, (approximate) parity and time-reversal invariance,
color gauge symmetry, baryon-number conservation,
and approximate $SU(2)_L \times SU(2)_R$ chiral symmetry. 
Although it is possible to extend the theory to incorporate
violations of parity and time-reversal,
we neglect them here
since they are small relative to the accuracy pursued by us.
Generalization to $SU(3)_L \times SU(3)_R$ is also possible.

Color gauge invariance is trivially satisfied in the EFT
because hadrons are color 
singlets. Baryon number conservation requires baryons to
appear in bilinears.
In the kinematic region where the EFT holds, external
momenta are much smaller than the nucleon mass, $Q \ll m_N$, and thus
Lorentz invariance can be fulfilled perturbatively in powers of
$Q/m_N$. 
There are two approaches in the literature to build the 
corresponding effective Lagrangian. One approach 
is to write a relativistic Lagrangian 
and then to derive the
$Q/m_N$ expansion by integrating out
high-energy components using the path integral \cite{Bernard:1992qa}. 
In the case of the delta \cite{hemmertdelta,pascalutsa},
one uses a Rarita-Schwinger field.
The so-called ``off-shell'' parameters that control the spurious spin-1/2
sectors of the Rarita-Schwinger field
are interpreted as choices of ``gauge''. By building a gauge-invariant
Lagrangian and choosing a certain gauge in Feynman rules, the spurious
spin-1/2 sectors can be removed from the final result.

It is not, however, inevitable
to rely on the form of a Lagrangian outside the regime of validity of an EFT;
only the symmetries should be important.
Since in the region where ChPT is valid the nucleon and
delta are always nonrelativistic,
the other approach \cite{jenkins,bira-thesis}
starts from the nonrelativistic limit.
This is accomplished by using heavy-fermion fields $N$ for
the nucleon and $\Delta$ for the delta, which are, respectively,
two- and four-component spinors in spin ($S$) and
isospin ($I$) spaces. 
Compared to the relativistic fields, the
heavy-fermion fields have the common, inert, large mass $m_N$
removed, and contain only the destruction of particles.
Not only does the
heavy-baryon formalism keep clear track of the small expansion
parameter, $Q/m_N$, but it also is convenient as a framework to write the
most general effective Lagrangian, where
the only baryon degrees of freedom are those representing
forward propagation.

As usual, we introduce the spin operators
$\vec{S}^{(S)}$, normalized so that
\beq
\big[ S^{(S)}_i, S^{(S)}_j \big] =
i\epsilon_{ijk} S^{(S)}_k\; ,
\eeq
and in isospace $\iv{t}^{(I)}$, normalized the same way.
We write $\vec{S}^{(\frac{1}{2})}=\vec{\sigma}/2$
and $\iv{t}^{(\frac{1}{2})}=\iv{\tau}/2$ in terms of Pauli
matrices.
One also needs 
transition matrices that have proper Clebsch-Gordan (CG)
coefficients embedded. 
We define $2 \times 4$ matrices $S_i$ in spin
space 
such that their matrix elements between a nucleon state with a 
spin $z$-component $\sigma$ and a delta state with a 
spin $z$-component $s$ are
\beq
\left( S_i \right)_{\sigma s} \equiv \langle 1 \frac{1}{2}\, ; \sigma \, i
| 1 \frac{1}{2} \,; \frac{3}{2}\, s \rangle \; , 
\eeq
where we use
the notation of Ref.~\cite{sakurai}. It is not difficult to show that the 
bilinear $N^{\dagger} \vec{S} \Delta$ is a three-vector, thanks to the
Wigner-Eckart theorem. We impose the normalization condition
\beq
S_i {S_j}^{\dagger} = \frac{1}{3} \left(2 \delta_{ij} - i
\epsilon_{ijk} \sigma_k \right)\; .
\eeq
In addition, there is a spin-2 
bilinear, $N^{\dagger}\Omega_{i j}\Delta$, which is a
symmetric, traceless three-tensor, with
\beq
\left( \Omega_{ij} \right)_{\sigma s} \equiv \langle 2 \frac{1}{2}\, ;
\sigma\, i\, j | 2 \frac{1}{2}\, ; \frac{3}{2}\, s \rangle 
= \frac{1}{2} \left( \sigma_i  S_j + \sigma_j S_i \right)_{\sigma s} \; .
\eeq
Similar transition matrices, $\iv{T}$ and $\Xi_{a b}$, can be
defined in isospace such that $N^{\dagger} \iv{T} \Delta$ and
$N^{\dagger} \Xi_{ab} \Delta$ are an isovector and an isotensor, 
respectively.

In order to build the effective Lagrangian, 
one first enumerates all the rotation-invariant operators and then uses a
Lorentz transformation rule perturbative in slow-velocity boosts to
constrain the coefficients of those operators. However, it is not yet
clear how one can go beyond order $Q/m_N$ with this ``bottom-up'' approach.
A separate paper by one of us \cite{bw-nonrel} addresses this issue:
the systematic construction of a deltaful, heavy-baryon chiral
Lagrangian with arbitrarily high relativistic corrections. 
The Appendix summarizes the aspects of Ref.~\cite{bw-nonrel} 
relevant for the current paper.
In essence, we implement baryons via
the Foldy-Wouthuysen representation \cite{FW50} of
the Poincar\'e group, whose boost generators can be readily
expanded in powers of $Q/m_N$.

Chiral symmetry is more
complicated to implement because it is spontaneously broken
by the QCD vacuum to its isospin subgroup, and it is thus
nonlinearly
realized in terms of pion and baryon fields 
\cite{wnbgnonliear,ccwz,weinbergbook}. 
Here we use stereographic coordinates $\iv{\pi}$ to represent
the isovector pion field \cite{wnbgnonliear,weinbergbook,bira-thesis}, 
for which we define
a covariant derivative
\beq
\iv{D}_{\mu} \equiv D^{-1} \frac{\partial_{\mu} \iv{\pi}}{2 f_\pi}  \; ,
\eeq
with $f_\pi \simeq 92$ MeV the pion decay constant and
\beq
D \equiv 1 + \frac{\iv{\pi}^2}{4 f_\pi^2} \; .
\eeq
The nonlinear realization maps
axial chiral-rotations of $N$, $\Delta$, and $\iv{D}_{\mu}$ into
$\iv{\pi}$-dependent (hence ``local'') isospin rotations 
\cite{wnbgnonliear,ccwz,weinbergbook}. 
Since such local isospin rotations do not commute
with normal derivatives, one also needs covariant derivatives for $N$,
$\Delta$, and $\iv{D}_\mu$. For a generic chiral-covariant field with
isospin $I$, $\psi^{(I)}$, its covariant derivative is defined as
\cite{wnbgnonliear,weinbergbook,bira-thesis}
\beq
\mathscr{D}_{\mu} \psi^{(I)} \equiv \left(\partial_{\mu} + \iv{t}^{(I)}
\bm{\cdot} \iv{E}_{\mu} \right) \psi^{(I)} \; ,
\eeq
where 
\beq\iv{E}_{\mu} \equiv i \frac{\iv{\pi}}{f_\pi} \bm{\times}
\iv{D}_{\mu} \; .
\eeq
For an isovector with Cartesian indices like $\iv{D}_\nu$, it is
conventional to write the covariant derivative as
\beq
\mathscr{D}_\mu \iv{D}_\nu \equiv \partial_{\mu} \iv{D}_{\nu} +
i \iv{E}_{\mu} \bm{\times} \iv{D}_{\nu} \; .
\eeq
Any isospin-invariant operator made of $N$, $\Delta$, 
$\iv{D}_\mu$ and their covariant derivatives will automatically be
$SU(2)_L \times SU(2)_R$ invariant.

Explicit chiral-symmetry breaking induced by the quark masses can
easily 
be incorporated in the effective Lagrangian. Those operators,
denoted by $\Phi^\pm$, that are proportional to 
$ m_u \pm m_d$ will have a structure 
as follows \cite{wnbgnonliear,weinbergbook,bira-thesis},
\bea
\Phi^+ & = & - D^{-1} \frac{\iv{\pi}}{f_\pi} \bm{\cdot} \iv{\eta}^+ +
D^{-1} \left( 1 - \frac{\iv{\pi}^2}{4 f_\pi^2} \right)
\eta^+_4 \; , \\
\Phi^- & = & \left( \eta^-_3 - \frac{1}{2} D^{-1} \frac{\pi_3}{f_\pi}
  \frac{\iv{\pi}}{f_\pi} \bm{\cdot} \iv{\eta}^- \right) 
+ D^{-1}\frac{\pi_3}{f_\pi} \eta^-_4 \; ,
\eea
where the quantities 
$\iv{\eta}^\pm$  and $\eta^\pm_4$ are built
of covariant fields and 
are isovector and isoscalar, respectively. To preserve parity, it is easy to
show that $\iv{\eta}^+$ ($\iv{\eta}^-$) and $\eta^-_4$ ($\eta^+_4$)
are pseudoscalar (scalar).
It can be shown \cite{bira-thesis} that
isospin is an accidental symmetry, in the sense that it only appears
in the subleading effective Lagrangian.
As a first
study, we focus here on the isospin-invariant part of the $\pi N$ amplitude.

Electromagnetic interactions can be easily incorporated
in the Lagrangian by adding the requirement of $U(1)$ gauge invariance.
This is accomplished by
turning all derivatives in existing interactions
into gauge-covariant
derivatives, and by adding additional gauge-invariant interactions
with the electromagnetic field strength. 
Here for simplicity we neglect electromagnetic interactions.

\subsection{Effective Lagrangian}

Since the effective Lagrangian has an infinite number of interactions,
one needs a scheme to organize all its operators.
It is convenient to order the
Lagrangian terms according to the so-called chiral index $\nu$
\cite{weinberg79}, 
\beq
\nu = d + m + n_{\delta} + \frac{f}{2} - 2 \ge 0 \; , \label{eqn:chiind}
\eeq
where $d$, $m$, $n_\delta$ and $f$ are the numbers of
derivatives, powers of $m_{\pi}$, powers of $\delta$ and fermion
fields, respectively. The lowest value of the index is a consequence
of the pattern of chiral-symmetry breaking in QCD.

In constructing the Lagrangian, we use 
integration by parts and field redefinitions to remove time derivatives on
baryon fields except for the kinetic terms. The 
Lagrangian terms
with the two lowest indices are given by \cite{bira-thesis}
\bea
\mathcal{L}^{(0)} & = & 2 f_\pi^2 \iv{D}^2
- \frac{m_{\pi}^2 }{2D}  \iv{\pi}^2 
+  N^{\dagger} i \mathscr{D}_0 N 
+ g_A N^{\dagger} \iv{\tau} \vec{\sigma} N \bm{\cdot} \cdot \vec{\iv{D}} 
\nonumber \\ 
& & + \Delta^{\dagger} \left( i \mathscr{D}_0 - \delta \right) \Delta
+ 4 g^{\Delta}_A \Delta^{\dagger} \iv{t}^{(\frac{3}{2})}
\vec{S}^{(\frac{3}{2})} \Delta \bm{\cdot} \cdot \vec{\iv{D}} 
+ h_A \left( N^{\dagger} \iv{T} \vec{S} \Delta+ H.c. \right) 
      \bm{\cdot} \cdot \vec{\iv{D}} 
+\cdots  \label{eqn:lag0} 
\eea
and
\bea
\mathcal{L}^{(1)} &=& 
N^{\dagger} \frac{\vec{\mathscr{D}}^2}{2 m_N}N
+ 2c\frac{m_\pi^2}{D} \iv{\pi}^2 N^{\dagger} N
+ \Delta^{\dagger} 
  \left[\frac{\vec{\mathscr{D}}^2}{2 m_N}-(c^\Delta-c)m_\pi^2\right]
  \Delta 
+ 2c^\Delta\frac{m_\pi^2}{D} \iv{\pi}^2 \Delta^{\dagger} \Delta
\nonumber\\
&&- \frac{h_A}{m_N} 
 \left( i N^{\dagger} \iv{T} \vec{S}\cdot\vec{\mathscr{D}}\Delta + H.c.\right) 
 \bm{\cdot}\iv{D}_0 + \cdots \; , \label{eqn:lag1}
\eea
while the next-higher index yields
\bea
\mathcal{L}^{(2)} &=& 
-
\frac{\Delta m_\pi^2}{2D^2}\iv{\pi}^2 -\frac{\delta}{2 m_N^2}
\Delta^{\dagger} \vec{\mathscr{D}}^2 \Delta 
+ \frac{h_A}{2 m_N^2} 
\left[
     \left(N^{\dagger} \iv{T} \vec{S}\vec{\mathscr{D}}^2 \Delta
   - 
N^{\dagger} \iv{T}\vec{S}\cdot\vec{\mathscr{D}}\vec{\mathscr{D}}\Delta\right)
     + H.c.   
\right] \bm{\cdot} \cdot\vec{\iv{D}} 
\nonumber \\
& &
+ \frac{h_A}{8 m_N^2} 
\left[
  \left(\delta_{lm} N^{\dagger} \iv{T}\vec{S}\cdot\vec{\mathscr{D}}\Delta 
 +3  N^{\dagger} \iv{T}S_l\mathscr{D}_m \Delta 
 -2i \epsilon_{ijl}  
N^{\dagger}\iv{T}\Omega_{im} \mathscr{D}_j\Delta  \right) + H.c.
\right]
  \bm{\cdot} \mathscr{D}_l\iv{D}_m 
\nonumber \\
& &
+ d\, \frac{m_\pi^2}{D} 
   \left(1 - \frac{\iv{\pi}^2}{4f_\pi^2}\right) 
\left( N^{\dagger}\iv{T}\vec{S}\Delta + H.c.\right)\bm{\cdot}\cdot\vec{\iv{D}}
+ \cdots 
\label{eqn:lag2}
\eea
Here, $g_A$ ($g^\Delta_A$) is the $\nu=0$ axial-vector coupling of the nucleon
(delta)
and $h_A$
($d$) is the $\nu=0$ ($\nu =2$) $\pi N \Delta$ 
coupling. 
These low-energy constants (LECs) are expected to be of 
$\mathcal{O}(1/\mqcd^\nu)$ but are not determined by chiral symmetry.
We define the phases of pion and delta fields so that 
$g_A\ge 0$ and $h_A\ge 0$.
The $\Delta m_\pi^2$ term provides a correction to the pion mass
that is proportional to the square of the average quark mass
(it is related to the $l_3$ term in Ref. \cite{weinberg79});
in the following, in order to simplify formulas, we absorb its
contribution in $m_\pi^2$.
Likewise, the nucleon and delta masses 
receive at $\nu=1$ contributions that are linear in the average quark mass,
the respective sigma terms (see, {\it e.g.}, Ref. \cite{anotherjenkins})
denoted here by $c$ and $c^\Delta$. 
With our choice of heavy-nucleon field we have already absorbed 
$cm_\pi^2$ in the nucleon mass $m_N$.
Again, for simplicity, in the following 
we absorb the remaining mass contribution, $(c^\Delta-c)m_\pi^2$,
in $\delta$.
The remaining pion-delta interaction, together
with a number of other interactions not shown, contributes 
to the order we work below only to a further renormalization of $\delta$.
The interactions
associated with the pion and nucleon mass corrections only contribute 
to our reaction at higher order.
Note that ``$\cdots$'' refer to terms that do not 
appear explicitly in $\pi N$ scattering to the order concerned in this paper.
Higher-index Lagrangians can be constructed with more derivatives,
but will also only contribute at higher orders.

Different versions of the heavy-baryon effective Lagrangian that are
deduced from a relativistic
formalism are given in Refs.~\cite{hemmertdelta,threshold-delta2}. 
Their Lagrangians, in
our notation, both have an independent $\nu=1$ $\pi N \Delta$ coupling
(denoted by $b_3$ and $b_3 + b_8$ respectively in
Refs.~\cite{hemmertdelta} and \cite{threshold-delta2}). 
This is because redundancy due to baryonic
equations of motion is only removed at the relativistic level in
Refs.~\cite{hemmertdelta, threshold-delta2}, and further minimization
of the number of interactions due to the heavy-baryon equations of motion 
is not
considered there. The $d$ term in Eq. \eqref{eqn:lag2} is equivalent to
the combination of couplings
$-2f_4+f_5$ in Ref.~\cite{threshold-delta2}.

\section{Power Counting\label{sec:powct}} 

We now turn to the ordering of contributions to physical processes.
For definiteness we take $\pi N$ elastic scattering, although 
the power counting is the same for other one-nucleon reactions
where the external CM energy can be dialed to near the delta-nucleon mass
splitting. We consider throughout the case $Q\sim m_\pi \sim \delta \ll \mqcd$.

\subsection{Away from the resonance}

We first consider CM energies much below the delta peak. 
In this case power counting is standard
\cite{weinberg79, BerKaiMei, bira99review} 
with the simple generalization that
$\delta$ counts as $Q$.
The contribution of a diagram with $A$ nucleons (here $A = 1$),
$L$ loops, and $V_i$ vertices
with chiral index $\nu_i$ is proportional to $Q^\rho$, with
\beq
\rho = 2 - A + 2L + \sum_i V_i \nu_i \ge 2 - A\, .
\label{eqn:std-con}
\eeq
The contributions with minimum $\rho$ form the leading order (LO), the next 
contributions are referred to as next-to-leading order (NLO), and so on.

\begin{figure}
\includegraphics[scale=0.8]{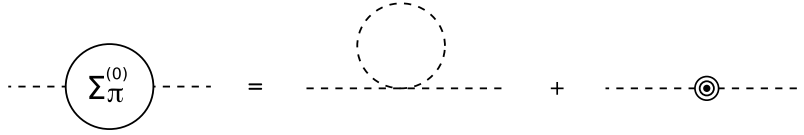}
\caption{\label{fig:pise}The LO pion self-energy, $\Sigma_\pi^{(0)}$.
A dashed line represents a free pion propagator. The unmarked vertex
has  $\nu=0$ and
the twice-circled vertex has $\nu = 2$.}
\end{figure}

The power counting can also be applied to sub-diagrams if $A$
is generalized to count any 
fermion line that is unattached on one side. 
Examples, which will be important later, are the following:

\noindent
{\it (i)} The LO contribution to the pion self-energy
$\Sigma_\pi^{(0)}=\mathcal{O}(Q^4/\mqcd^2)$, see \reffig{fig:pise},
two powers down compared to the inverse of the free pion propagator.
{}From the LO pion self-energy we obtain a correction to the
pion-field renormalization constant, $Z_\pi^{(2)}=\mathcal{O}(Q^2/\mqcd^2)$.

\begin{figure}
\includegraphics[scale=0.8]{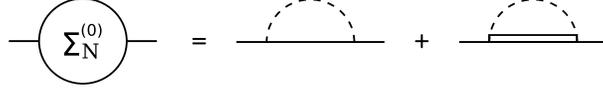}
\caption{\label{fig:nse}The LO nucleon self-energy, $\Sigma_N^{(0)}$.
A single (double) line represents a free nucleon (delta)
propagator. Vanishing diagrams are not shown.}
\end{figure}

\noindent
{\it (ii)} The LO contribution to the nucleon self-energy
$\Sigma_N^{(0)}=\mathcal{O}(Q^3/\mqcd^2)$, see \reffig{fig:nse},
two powers down from the nucleon kinetic energy.
{}From the LO nucleon self-energy we obtain a correction to the
nucleon-field renormalization constant,
$Z_N^{(2)}=\mathcal{O}(Q^2/\mqcd^2)$. 

\begin{figure}
\centering
\includegraphics[scale=0.9]{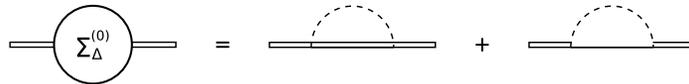}
\caption{\label{fig:se0} The LO delta self-energy,
  $\Sigma_\Delta^{(0)}$. Vanishing diagrams are not shown.}
\end{figure}

\noindent
{\it (iii)} The LO contribution to the delta self-energy
$\Sigma_\Delta^{(0)}=\mathcal{O}(Q^3/\mqcd^2)$, see \reffig{fig:se0},
two powers down from the delta kinetic energy and delta-nucleon
mass difference. 
{}From the LO delta self-energy we can define a correction to the
delta-field renormalization constant, 
$Z_\Delta^{(2)}=\mathcal{O}(Q^2/\mqcd^2)$.

\begin{figure}
\centering
\includegraphics[scale=0.8]{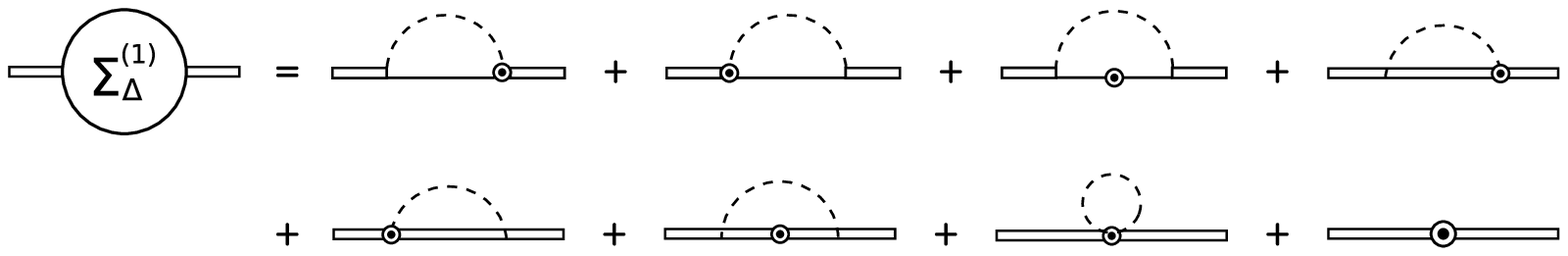}
\caption[NLO delta self-energy]{\label{fig:se1} The NLO correction
  to the delta self-energy, $\Sigma_\Delta^{(1)}$.
The once-circled vertices have $\nu= 1$. Vanishing diagrams are not shown.}
\end{figure}

\noindent
{\it (iv)} The NLO contribution to the delta 
self-energy
$\Sigma_\Delta^{(1)}=\mathcal{O}(Q^4/\mqcd^3)$, see \reffig{fig:se1}.

\begin{figure}
\centering
\includegraphics[scale=0.9]{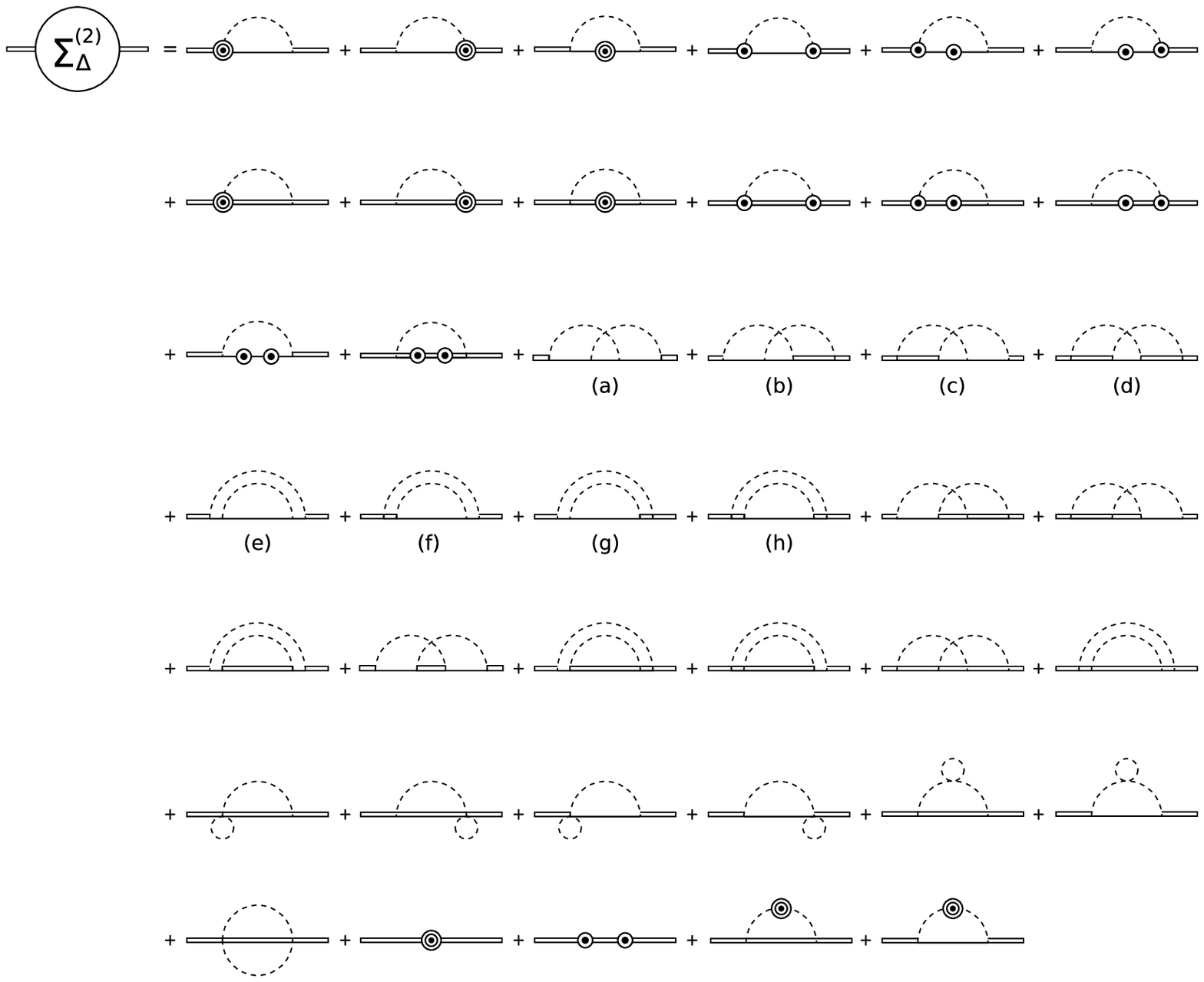}
\caption[NNLO delta self-energy]{\label{fig:se2} The NNLO
  correction to the delta self-energy, $\Sigma_\Delta^{(2)}$.
The twice-circled vertices have $\nu = 2$. Vanishing diagrams are not shown.}
\end{figure}

\noindent
{\it (v)} The NNLO contribution to the 
delta self-energy
$\Sigma_\Delta^{(2)}=\mathcal{O}(Q^5/\mqcd^4)$, see \reffig{fig:se2}.

\begin{figure}
\centering
\includegraphics[scale=0.75]{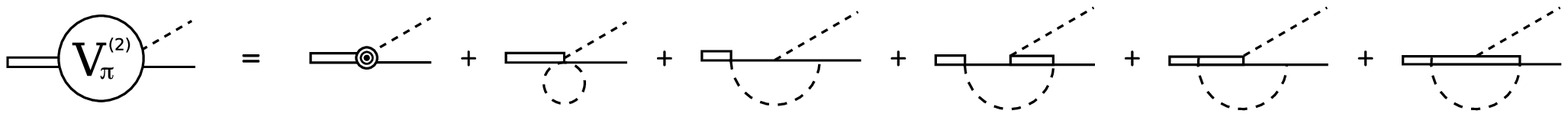}
\caption[NNLO $\pi N \Delta$ vertex]{\label{fig:f2}
The NNLO correction to the $\pi N \Delta$ vertex, $V_\pi^{(2)}$. 
Vanishing diagrams are not shown.} 
\end{figure}

\noindent
{\it (vi)} The NNLO contribution to the $\pi N \Delta$ vertex form-factor
$V_\pi^{(2)}=\mathcal{O}(Q^3/\mqcd^2)$, see \reffig{fig:f2},
two powers down from the $\nu=0$ $h_A$ vertex.

This power counting holds for generic momenta,
but it does not work equally well in every specific region of phase space.
For example, close to threshold an incoming or outgoing pion has energy 
very close to $m_\pi$ and a three-momentum close to zero ---it is, in
other terms, nonrelativistic.
In such cases, a spatial derivative on the pion
field does not contribute the same as
a time derivative, yet Eq.~\eqref{eqn:std-con} does not discriminate
between these derivatives. One can refine power counting
for this region if one wishes.

\subsection{Near the resonance}

\begin{figure}
\centering
\includegraphics[scale=1]{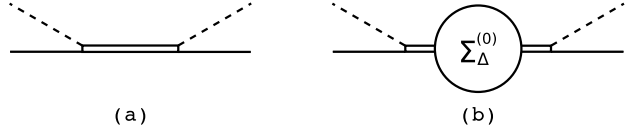}
\caption[Example of one-$\Delta$-reducible diagram]{\label{fig:egkft} 
Examples of one-$\Delta$-reducible diagrams.} 
\end{figure}

It should thus be no surprise that the above power counting fails in the
immediate neighborhood of the delta resonance.
Consider the two contributions to $\pi N$ scattering in
\reffig{fig:egkft} at a CM energy $E$. 
Diagram (a) is proportional to $1/(E - \delta)$
and diagram (b) to $\Sigma_\Delta^{(0)}(E)/(E - \delta)^2$,
which are, respectively, $\mathcal{O}(1/Q)$ and $\mathcal{O}(Q/M_{\rm QCD}^2)$
at a generic low energy.
Although both $E$ and $\delta$ are separately of $\mathcal{O}(Q)$, they have
opposite signs and can cancel. 
When that happens,
diagrams (a) and (b) are enhanced above their standard power counting,
and diagram (b) is enhanced more than (a) so that 
the self-energy might
no longer be a small correction: a resummation is necessary.
The situation here is completely analogous to other 
narrow resonances \cite{resonances},
where a ``kinematic fine-tuning'' requires a modification of 
power counting in the neighborhood of a resonance.

\begin{figure}
\centering
\includegraphics[scale=1]{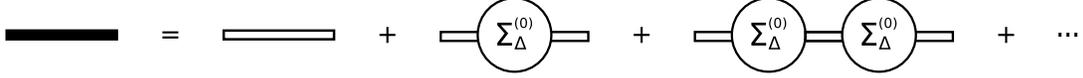}
\caption[Dressed $\Delta$ propagator at LO]{\label{fig:bubblesum}
  Dressed delta propagator at LO as a sum of insertions of
  $\Sigma_\Delta^{(0)}$.}
\end{figure}

In fact, within a window of size 
\beq
|E - \delta| = \mathcal{O}\left(\frac{Q^3}{\mqcd^2}\right)
\eeq
around the delta peak, the bare delta propagator is 
$\mathcal{O}(\mqcd^2/Q^3)$. 
Since $\Sigma_\Delta^{(0)}$ is $\mathcal{O}(Q^3/\mqcd^2)$,
one simultaneous insertion of $\Sigma_\Delta^{(0)}$ and the bare delta 
propagator
contributes $\mathcal{O}(1)$, so the two diagrams in \reffig{fig:egkft}
become comparable.
We should thus resum 
the geometric series of one-$\Delta$-reducible delta
propagators shown in  \reffig{fig:bubblesum}.
Moreover, since within the window $E=\delta (1+ \mathcal{O}(Q^2/\mqcd^2))$,
the energy dependence of $\Sigma_\Delta(E)$ is always two powers down from
its value at $E=\delta$:
\beq
\Sigma_\Delta^{(n)}(E) = 
\Sigma_\Delta^{(n)}(\delta)\left\{1 + \mathcal{O}\left(\frac{Q^2}{\mqcd^2}
\right)\right\}\; .
\label{eqn:seexp}
\eeq
The resummation thus amounts to a dressed propagator
\bea
S_\Delta^{(0)}(E)= \left[E-\delta+\Sigma_\Delta^{(0)}(\delta)\right]^{-1},
\label{eqn:dressedp}
\eea
which scales as $\mqcd^2/Q^3$.
This is an enhancement of two powers over the generic situation.
To make a full amplitude, one needs to contract the dressed propagators
with the $\pi N \Delta$ vertices, for which we
should also neglect the energy dependence:
\beq
V_\pi^{(n)}(E) = 
V_\pi^{(n)}(\delta)\left\{1 + \mathcal{O}\left(\frac{Q^2}{\mqcd^2}
\right)\right\}\; .
\eeq

In contrast, in one-$\Delta$-irreducible diagrams the delta propagators
are far away from their pole and do not need to be dressed,
continuing to scale as $1/Q$.
This is trivial in one-$\Delta$-irreducible trees, \textit{e.g.}, the
$u$-channel $\Delta$-exchange diagram. Less trivial are the loop diagrams
where there are integrations over the energy domain spanning the delta pole. 
In this case, pions in the loops carry at least $m_\pi$
of energy, and the delta will not go on-shell for $\pi N$ energies
below $\delta +m_\pi$. Even at this point, the delta will be recoiling
and the kinetic energy needs to be taken into account before the self-energy.

Take for example the one-loop diagrams with a delta that contribute
to the LO nucleon (\reffig{fig:nse}) and delta (\reffig{fig:se0}) 
self-energies.
Suppose that
the external fermion and the internal
pion have four-momentum $p$ and $l$, respectively. Apart from the
CG coefficients and coupling constants in front, the loop integral 
with the bare dressed propagator has the form
\beq
\int \frac{d^3l}{(2\pi)^3} \int \frac{dl_0}{2\pi} 
  \frac{\vec{l}^{\;2}}{p_0 - l_0 - \delta +i\epsilon} 
  \frac{1}{l^2 - m_\pi^2 + i\epsilon} \; .
\eeq
One can always close the contour of integration over $l_0$ in the half-plane
opposite to the half-plane where the pole of the delta propagator is.
In this case, we pick the pole at $l_0=\omega-i\epsilon$,
where $\omega=\sqrt{\vec{l}^{\; 2}+m_\pi^2}$ is the pion energy.
The remaining integral over $\vec{l}$ has no singularities when $p_0=\delta$.
Alternatively, rewriting it as an integral over $\omega$ after
carrying out the angular integrations, it starts at $m_\pi$ 
while the integrand has poles only at $\omega=0$ and $\omega=p_0-\delta$. 
The same argument holds for $A=1$ diagrams with more loops.

As a consequence, 
in one-delta-irreducible diagrams the standard ChPT power counting 
\eqref{eqn:std-con}
still applies;
dressed delta propagators only need to be included in 
one-delta-reducible diagrams.
We thus arrive at a new power counting for one-$\Delta$-reducible
diagrams within a narrow window around the delta peak,
\beq
\rho = 2 - A - 2 n_\Delta + 2L + \sum_i V_i \nu_i 
\ge 2 - A - 2 n_\Delta \, ,
\label{eqn:rho2}
\eeq
where $n_\Delta$ is the number of dressed delta propagators. 
This is the non-electromagnetic version of $\rho$ derived
in a slightly different power counting in Ref. \cite{pascalutsa}.
We discuss the similarities and differences 
between the two approaches in Sec.~\ref{sec:other}.

Notice that there is a larger region around the resonance,
of size $|E - \delta| = \mathcal{O}(Q^2/\mqcd)$, where the
enhancement in the delta propagator is insufficient to 
compensate for an insertion of the self-energy.
Although the power counting is
slightly different than Eq. \eqref{eqn:std-con},
this case is still perturbative. For simplicity, we do not consider
it separately from the generic situation away from the resonance,
where $|E - \delta| = \mathcal{O}(Q)$.

\subsection{Sewing the two regions\label{sec:sewing}}

\begin{figure}
\centering
\includegraphics[scale=1.3]{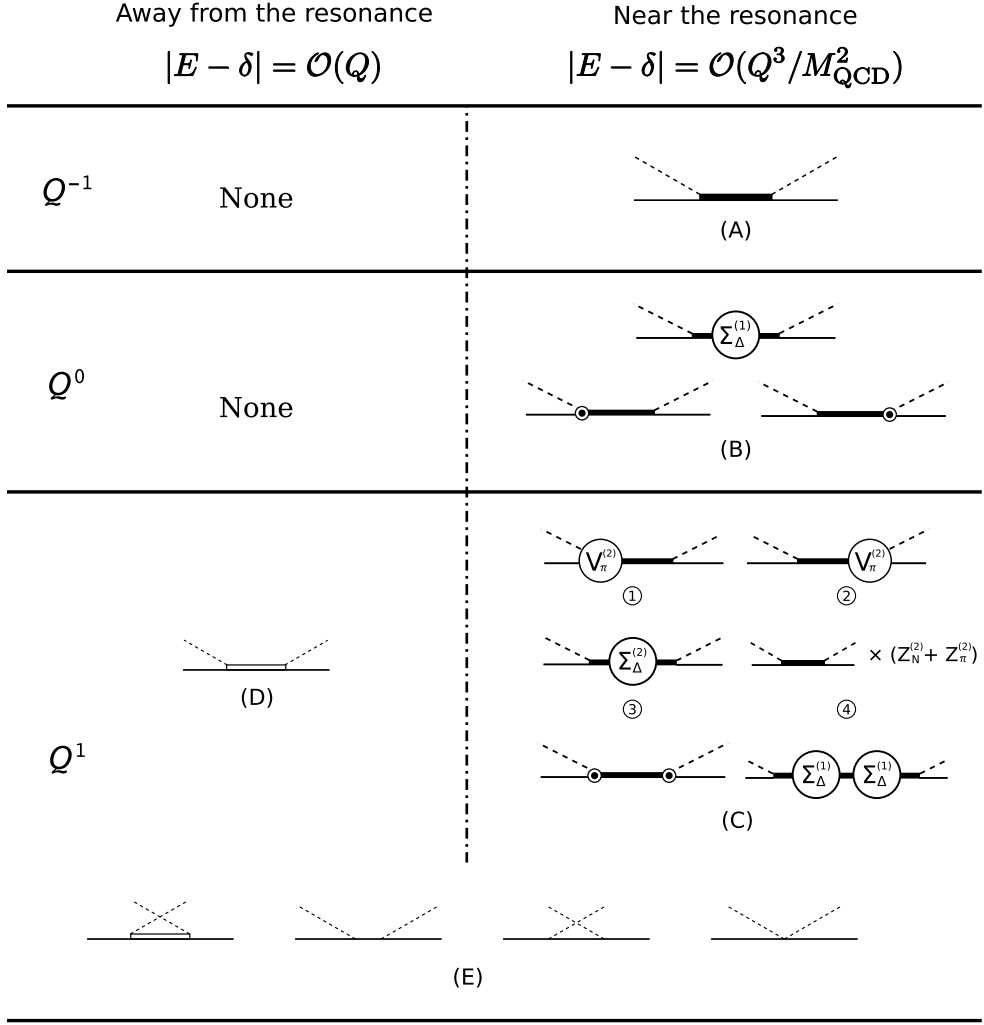}
\caption{\label{fig:powct}Contributions to $\pi N$ scattering 
up to order $Q^1$: 
(A) $Q^{-1}$ pole diagram; 
(B) $Q^{0}$ pole diagrams;
(C) $Q^{1}$ pole diagrams;
(D)\&(E) $Q^1$ tree diagrams,
of which (E) apply to both regions. 
Power counting away from the resonance is the standard ChPT counting. 
The relation between the double line (bare delta propagator)
and the thick solid line (dressed delta propagator) is given in 
Fig. \ref{fig:bubblesum}.
}
\end{figure}

We can now weigh the diagrams contributing to $\pi N$ scattering,
putting together the two power countings
\eqref{eqn:std-con} in the off-the-pole region
and \eqref{eqn:rho2} in the pole region.
These contributions generally scale as 
$\mathcal{O}(Q^\rho/\mqcd^{\rho-1}f_\pi^2)$, where
$1/f_\pi^2$ is due to the two pion external legs. Without causing
confusion, we simply specify the size of the contribution of a $\pi N$ 
scattering diagram by order ``$Q^\rho$''. Found in
\reffig{fig:powct} are diagrams up to order $Q^1$. 
To this order, the kinematic fine-tuning to the delta
simply promotes the diagrams on the right (labeled A, B, C) 
of the figure with respect to
the standard assignment on the left (D).
The diagrams at the bottom (E), which are one-delta irreducible,
are not affected.

It seems that the two different power-counting schemes, which are
applicable in two different regions, would lead to an EFT amplitude in
the form of a piecewise function in the energy. Even worse, separating
these two regions is somewhat arbitrary.

A piecewise EFT is actually unnecessary.
Look at, for instance, the $Q^{-1}$ order. Only the LO pole diagram
(\reffig{fig:powct}(A)) contributes at this order.
Since there is no diagram in the off-the-pole region,
the EFT prediction should vanish when away from the
pole. Enforcing the LO pole
diagram in the off-the-pole region seems to make a ``wrong'' prediction. 
However, it is only wrong by an amount of order $Q^1$,
since a dressed delta propagator scales off-pole as a
bare one ($E - \delta$ dominates over $\Sigma_\Delta^{(0)}$).
This is of the same order as the tree in \reffig{fig:powct}(D).
On the other hand, however, one already made an even larger error of
order $Q^0$ by neglecting the NLO pole diagrams. To put it another
way, extending the domain of the LO pole diagram is 
equivalent to shifting a subset of higher diagrams into the order
$Q^{-1}$. This does not disobey the original power counting as long as
one does not claim a higher accuracy
by doing so. In many perturbative quantum field theories,
one runs the renormalization group to select an optimal
renormalization scale in favor of a more rapid convergence of the
perturbative series, which is
also equivalent to a rearrangement of diagrams. 

This sort of rearrangement, written symbolically as
\beq
\text{N}^\alpha\text{LO} + \text{a subset of}\; \sum_{i > \alpha}
\text{N}^i\text{LO} \to \text{N}^\alpha\text{LO} \; ,
\label{eqn:rarule}
\eeq
must be used with caution. First, 
the 
higher-order subset being added to $\mathrm{N^{\alpha}LO}$ should be
cutoff-independent by itself, in order
to avoid introducing model dependence in the form of cutoff dependence.
Second, one probably would not like to move
undetermined LECs into lower orders because
doing so weakens the predictive power of EFT.
In our case, order-by-order renormalizability 
of the rearranged diagrams
will be relatively simple to demonstrate because,
in the resonance counting, energy variation in the delta self-energy
and interaction vertices is treated perturbatively.
We will also see that the rearranged diagrams
include in the off-pole region, within an error of order $Q^2$, only
delta parameters that appear in the diagrams of \reffig{fig:powct}(D,E).

Ensuring that our amplitude is correct also away from
the resonance region
does require some care.
When the pion and nucleon in the initial and final states are
on-shell, the momentum dependences in
the $\pi N \Delta$ external vertices 
translate into energy dependence. 
In the resonance counting,
such energy dependence is subject to $E = \delta +
\mathcal{O}(Q^3/\mqcd^2)$ so that the $\pi N \Delta$ external vertices 
in the LO and NLO pole diagrams are constant. This has two consequences. 
\textit{(i)} Up to order $Q^{0}$ the $P$-wave amplitude in the delta 
channel does not vanish at threshold as it should. 
But, as we argued, applying the LO pole diagram near
threshold is expected to make an error of order $Q^1$, which can be
taken as ``vanishing'' in 
comparison with the dominant amplitude around the resonance. 
\textit{(ii)} Since the energy
expansion around $\delta$ should not be enforced in \reffig{fig:powct}(D),
simply continuing the LO pole diagram away from the resonance would not 
reproduce \reffig{fig:powct}(D) 
due to the lack of energy
dependence in the external vertices. 
To account for
\reffig{fig:powct}(D), the easiest way is 
to restore at order $Q^{1}$ the energy dependence of
the external vertices of \reffig{fig:powct}(A).

To summarize, with sufficient caution 
it is unnecessary to restrict the energy domain of the
dressed pole diagrams in \reffig{fig:powct}.
Of course, our power counting stresses the fact that
amplitudes are much larger in the resonance region
than at threshold.
If one is interested solely in the region near threshold, where
the delta enhancement is not relevant, one is better off with
the standard ChPT power counting
---or, for reactions where typical three-momenta are below the pion mass,
with a ``heavy-pion'' EFT \cite{beeingsavage}.
Using our power counting in both threshold
and resonance regions is only efficient
if one aims for a unified description throughout the low-energy
region.

\subsection{Other approaches}
\label{sec:other}

The above power counting is not limited to $\pi N$ scattering.
It applies to any reaction where one can dial the initial energy
to bring the delta close to being on-shell. Of course, 
the need to account for the delta self-energy has been felt
for a very long time, and simple tree-level models with an added
delta self-energy already provide at least qualitative descriptions
of data \cite{ericson}.
What EFT provides in addition is a way,
consistent with QCD symmetries, to correct systematically for 
quantum effects and physics at short distances.
We now compare our power
counting to other approaches
that considered
non-perturbative effects of the delta in EFT. 

Reference \cite{ellis-tang} resummed the LO delta self-energy
in one-delta-reducible diagrams (as done here), and used the corresponding
delta width as an independent empirical input.
The scheme is quite different
from ours. First,
it did not take into account the fact
that the dressed delta propagator is enhanced by $\mathcal{O}(\mqcd^2/Q^2)$ 
in the resonance region, and
hence adopted the standard ChPT power counting except for simply
substituting the LO dressed delta
propagator for the bare one.
Therefore, the off-the-pole region
was excessively emphasized with a two-order-higher accuracy than the resonance
region. This discrepancy in power counting may be viewed as 
a rearrangement of the type \eqref{eqn:rarule}
but Ref.~\cite{ellis-tang} 
enlisted many more LECs to achieve the same overall
accuracy. Second, perhaps more importantly, 
Ref.~\cite{ellis-tang} 
did not consider the corrections to the delta
self-energy. For instance, in our power counting the one-loop
corrections to the $\pi N \Delta$ vertex, 
$V_\pi^{(2)}$ defined in \reffig{fig:f2},
contribute to the NNLO pole diagrams not only through vertices
attached to external legs but also in $\Sigma_\Delta^{(2)}$. The latter,
however, were neglected in Ref.~\cite{ellis-tang}. This would make one, 
if using the scheme in Ref.~\cite{ellis-tang},
unable to recover \refeq{eqn:bwback} around the resonance, which is
required by order-by-order unitarity. The difficulty of preserving
unitarity around the resonance forced 
Ref.~\cite{ellis-tang} to rely on certain
prescriptions, referred to as $S$- or $K$-matrix method therein, to
extract the phase shifts from the scattering amplitude. In particular,
one of the prescriptions, the $S$-matrix method, leads to a
discontinuity in the phase shifts.

More sophisticated methods of unitarization based on EFT exist,
for example Refs. \cite{unitarization1, unitarization2}.
In this case, if a sufficiently high-order kernel is fully iterated
in an analogous fashion to what is done in the two-nucleon system 
\cite{original,bira-thesis,bira99review},
corrections to the leading delta self-energy are accounted for
and good results can be obtained even beyond the delta resonance.
However, in a system with two heavy particles of reduced mass $\mu$
such a unitarization is justified by an infrared enhancement
of ${\cal O}(4\pi \mu/Q)$ over the standard ChPT counting 
\cite{bira99review}.
While this might apply to the strangeness sector of ChPT,
the strict EFT rationale for
resummation in pion-nucleon scattering in the resonance
region
must be rooted in arguments
such as those in Refs.~\cite{pascalutsa,morepascalutsa1,morepascalutsa2}
and in the present manuscript.
Contrary to Ref.~\cite{ellis-tang}, the resummations in
Refs.~\cite{pascalutsa,morepascalutsa1,morepascalutsa2} 
are based on a power counting very similar to ours.
Both in Refs.~\cite{pascalutsa,morepascalutsa1,morepascalutsa2} 
and here, it is recognized that the immediate vicinity of the resonance
requires a different power counting than the one in standard ChPT.
Moreover, in both approaches there is an attempt to smoothly
bridge the resonance region with lower energies.

Our power counting differs from the one proposed in
Ref.~\cite{pascalutsa}, however, in that 
$\delta$ was assumed
in Ref.~\cite{pascalutsa} to be much larger than
$m_\pi$, $m_\pi/\mqcd \sim (\delta/\mqcd)^2$.
As a consequence, the relative importance of explicit
chiral-symmetry-breaking terms is reduced.
Strictly speaking, in the light of the hierarchy 
$m_\pi \ll \delta$,
one would have to neglect the $m_\pi^2$ in
the pion propagator in $\Sigma_\Delta^{(0)}$ and 
other diagrams where the pion momentum is
$\mathcal{O}(\delta)$. This might lead to unpleasant infrared divergences in
certain diagrams, which Ref.~\cite{pascalutsa} avoids by not enforcing
their power counting in the calculation of diagrams. Of course this
can be justified by a rearrangement in the fashion of
\refeq{eqn:rarule}, and if this is done it blurs
the difference between the two power
countings somewhat.

Clearly, as emphasized by the authors,
the power counting of Ref.~\cite{pascalutsa} is well-suited
to study the regime of smaller quark masses, where 
$m_\pi$ decreases but $\delta$ is basically unchanged.
In contrast, it works less well as the number of colors $N_c$ increases, 
when $\delta$ decreases.
In any case, in the real world the scales are not clearly
separated, $\delta$ being larger than
$m_\pi$ only by a factor of $\sim 2$, and 
the interesting limits
$m_\pi\to 0$ and $N_c\to \infty$ can always be studied separately
afterwards.

Thus, if one is going to generically treat the pion mass as comparable
to momenta in loops around the resonance,
we find it simpler not to emphasize such a factor of 2, given that 
there are lots of other similar factors floating around. 
We simply count $\delta$ as comparable to the $m_\pi$, 
$m_\pi/\mqcd \sim \delta/\mqcd$, as it has been done 
before away from the resonance 
\cite{jenkins, hemmertdelta,threshold-delta1,threshold-delta2,bira-thesis,
kaiserbrock}.

This discussion about the best way to power-count explicit-symmetry-breaking 
terms should not obscure ---as it apparently has---
the important fact that it is the kinematic fine-tuning to a narrow
resonance (that is, one that has a width smaller than its energy)
that demands a resummation \cite{pascalutsa}.
This is in fact quite a general requirement 
of an EFT for shallow resonances \cite{resonances},
which has nothing do to with explicit chiral-symmetry breaking. 

\section{One-delta-irreducible ingredients\label{sec:pre}}

As an example of our approach, we want to calculate the $\pi N$
$T$ matrix to $\mathcal{O}(Q^1)$.
The tree diagrams are straightforward, but 
the pole diagrams in \reffig{fig:powct} 
are more complicated. The latter have three common
components:  
the field renormalization constants $Z_\pi$ and $Z_N$,
the $\pi N \Delta$ vertex function $V_\pi$,
and the delta self-energy $\Sigma_\Delta$.
All of these ingredients are 
made up of one-$\Delta$-irreducible graphs, which 
can be expanded in powers of $Q/\mqcd$ according to the standard
ChPT power counting. 
In this section we investigate each of the ingredients in turn,
up to relative $\mathcal{O}(Q^2/\mqcd^2)$.

Note that in general these quantities are cutoff-dependent.
If dimensional regularization (DR) is used, cutoff
dependences in the loops appear as  $1/(D-4)$ poles
in the complex dimensionality ($D$) plane near $D = 4$. Of course, the
scattering amplitude should be independent of any cutoff. 
Therefore, the cutoff
dependences must be absorbed by suitable counterterms.

\subsection{\label{sec:frc}Field renormalization constants}

Field renormalization constants $Z_\pi$ and $Z_N$ are
the residues of the  corresponding two-point Green's function (or
fully dressed propagator). 

The renormalized pion mass is the magnitude of the momentum of the pole of the 
dressed pion propagator, which we denote by $m_\pi$. At the pole,
the pion four-momentum $p$ obeys $p_0 = \omega(|\vec{p}|)$,
the energy of an on-shell pion:
\beq
\omega(|\vec{p}|)\equiv \sqrt{\vec{p}^{\, 2}+m_\pi^2} \, .
\label{eqn:deltak} 
\eeq
The pion self-energy, $\Sigma_\pi(p^2)$, is such that $\Sigma_\pi(m_\pi^2)=0$.
The pion field renormalization constant $Z_\pi$ is
related to $\Sigma_\pi(p^2)$ by
\beq
Z_\pi^{-1} = 1 + \frac{d}{d p^2}\Sigma_\pi(p^2)\bigg{|}_{p^2=m_\pi^2}\; .
\eeq
Expanding 
in powers of $Q/\mqcd$ according to
the standard ChPT power counting, to NNLO,
\bea
Z_\pi^{(0)} &=& 1\; ,\label{Zpiexp0}\\
Z_\pi^{(1)} &=& 0\; ,\label{Zpiexp1}\\
Z_\pi^{(2)} &=& -\frac{d}{d p^2}\Sigma_\pi^{(0)}(p^2)\bigg{|}_{p^2 =
  m_\pi^2}\; ,
\label{Zpiexp}
\eea
where $\Sigma_{\pi}^{(0)}$ 
is the LO pion self-energy shown in \reffig{fig:pise}.

Similarly, we renormalize the nucleon mass in such a way that the
pole of the dressed propagator of a nucleon
of four-momentum $p$ is given by 
$p_0=E_N(|\vec{p}|)$,
where 
\beq
E_N(|\vec{p}|) \equiv \sqrt{\vec{p}^{\, 2} + m_N^2} - m_N
\label{EN}
\eeq
is the heavy-baryon energy of an on-shell nucleon.
The nucleon self-energy $\Sigma_N(p)$
satisfies $\Sigma_N(E_N(|\vec{p}|), \vec{p})=0$,
and $Z_N$ is related to $\Sigma_N(p)$ by
\beq
Z_N^{-1} = 1 + \frac{\partial}{\partial p_0}\Sigma_N(p)\bigg{|}_{p_0 =
  E_N(|\vec{p}|)}\; .
\eeq
To NNLO,
\bea
Z_N^{(0)} &=& 1\; ,\label{ZNexp0}\\
Z_N^{(1)} &=& 0\; ,\label{ZNexp1}\\
Z_N^{(2)} &=& -\frac{\partial}{\partial p_0}\Sigma_N^{(0)}(p)\bigg{|}_{p_0 =
  0}\; ,
\label{ZNexp}
\eea
where $\Sigma_N^{(0)}$ 
is the LO 
nucleon self-energy shown in \reffig{fig:nse}.

The field renormalization constants are not directly observable.
Not surprisingly, $Z_N^{(2)}$ and $Z_\pi^{(2)}$ are cutoff-dependent.

\subsection{\label{sec:Dprop}$\Delta$ propagator}

The loops appearing in the delta self-energy,
$\Sigma_\Delta(p)$, have both real and
imaginary components. The inverse delta propagator is written as
\beq
S_\Delta^{-1} = 
p_0 - E_\Delta(|\vec{p}|)
     +  \textrm{Re}\Sigma_\Delta(p) + \frac{i}{2}\gamma(p) \; ,
\eeq
where
\beq
E_\Delta(|\vec{p}|) \equiv \sqrt{\vec{p}^{\, 2}+m_\Delta^2} - m_N
\label{ED}
\eeq
is the heavy-baryon energy of the delta, and
\beq
\gamma(p) \equiv 2\text{Im}\Sigma_\Delta(p)\; .
\label{eqn:defgamma}
\eeq
We choose to renormalize $m_\Delta$
so that 
\beq
\textrm{Re}\Sigma_\Delta(E_\Delta(|\vec{p}|),\vec{p}) = 0 \, .
\label{eqn:deltarenorm}
\eeq
The regulator dependence that arises in
the real part of the self-energy $\Sigma_\Delta(p)$ at the on-shell
point, $E_\Delta(|\vec{p}|)$, is thus absorbed in $\delta$.

Although the delta need not be an asymptotic state at least in
low-energy $\pi N$ elastic scattering, it is useful to introduce
the field renormalization
constant $Z_\Delta$, which we \emph{define} as
\beq
Z_\Delta^{-1} \equiv 1+ \frac{\partial}{\partial p_0} 
\left[\textrm{Re}\Sigma_\Delta(p)\right]\bigg{|}_{p_0 =E_\Delta(|\vec{p}|)}
\; .
\eeq
Since the energy dependence of $\Sigma_\Delta(p)$ only appears at NNLO,
the leading few orders of $Z_\Delta$ are given by
\bea
Z_\Delta^{(0)} &=& 1\; , \label{eqn:zd0}
\\
Z_\Delta^{(1)} &=& 0\; , \label{eqn:zd1}
\\
Z_\Delta^{(2)} &=& -\frac{\partial}{\partial p_0}
\left[\textrm{Re}\Sigma_\Delta^{(0)}(p)\right]\bigg{|}_{p_0 = \delta}\;, 
\label{eqn:zd2}
\eea
where $\Sigma_\Delta^{(0)}$ is the LO delta self-energy given 
in \reffig{fig:se0}.
We stress that $Z_\Delta$ is merely an intermediate quantity in our
calculation and the definition above is not unique by any means. As we will
see in Sec.~\ref{sec:t}, the amplitude in the end does not depend explicitly
on $Z_\Delta$.

\subsection{$\pi N \Delta$ vertex}

We \emph{define} the $\pi N \Delta$ vertex function, $V_\pi$, as the sum of all
\emph{amputated} $\Delta \to \pi N$ subdiagrams that have an 
incoming delta carrying four-momentum $(p_0, \vec{p})$
and an outgoing pion with four-momentum $(k_0, \vec{k})$ and isospin $a$. The
incoming and outgoing particles are not
necessarily on-shell. Rotational and isospin invariance require that
$V_{\pi\,a}$ be a $2\times4$ matrices in spin and isospin space of 
the form
\beq
V_{\pi\,a} =
T_a \left[ \vec{S}\cdot\vec{k}\, F + \vec{S}\cdot\vec{p}\,
  G + \epsilon_{ijl}\Omega_{im}k_j p_l\left(k_m H + p_m Q\right)
\right]\; ,
\eeq
where $F$, $G$, $H$, and $Q$ are three-scalar form factors that depend on
$p_0$, $k_0$, $\vec{p}\,^2$, $\vec{p}\cdot\vec{k}$, and $\vec{k}\,^2$.
For simplicity, we study only the special case where the incoming
delta sits in its own CM frame, \textit{i.e.}, $\vec{p}=0$. 
The vertex function is simplified as
\beq
V_{\pi\,a}^{\tr{CM}} = T_a \vec{S}\cdot\vec{k} \, F(p_0, k_0, k^2)\; ,
\eeq
with $k \equiv |\vec{k}|$.

The function $F$ at LO and NLO,
$F^{(0)}$ and $F^{(1)}$ respectively, 
is constant and can be read off directly from
$\mathcal{L}^{(0)}$ \eqref{eqn:lag0} and $\mathcal{L}^{(1)}$ \eqref{eqn:lag1}.
At NNLO, $F^{(2)}(p_0, k_0, k^2)$ consists of $\nu =2$ 
interactions and one-loop diagrams, shown in
\reffig{fig:f2}.
$F^{(2)}(p_0, k_0, k^2)$ has cutoff dependences proportional to
$p_0^2$, $k_0^2$, $p_0k_0$, $\delta p_0$, $\delta k_0$, $\delta^2$,
and $m_\pi^2$. 
Naively, one could introduce $\nu=2$ counterterms that
are associated with
time derivatives on $\Delta$ and/or $\iv{\pi}$,
in order to absorb divergences associated with $p_0$ and $k_0$. 
However, those counterterms can always 
be removed using equations of motion or field redefinitions, that is,
they are redundant
parameters, which $S$-matrix elements do not depend upon.
Fortunately, the $\pi N \Delta$ vertex function need not be
cutoff-independent. Since
there are other cutoff dependences floating around, such as those in
the field renormalization constants and self-energies,
what matters is that the combined
cutoff dependence cancels out when the ingredients are put
together in the EFT amplitude. 

As we will see in Sec.~\ref{sec:amps}, a
cutoff-independent form factor, $F_R$, will appear in the amplitude,
\beq
F_R \equiv \sqrt{Z_\pi} \sqrt{Z_N} \sqrt{Z_\Delta}\, 
F(\delta,\omega( k_\delta), k_\delta^2)\; , \label{eqn:defFR}
\eeq
where 
$k_\delta$ is the momentum of the pion when all particles are on-shell,
\beq
k_\delta = \left(\delta^2-m_\pi^2\right)^\frac{1}{2}
    \frac{\left[1+\delta/m_N+(\delta^2-m_\pi^2)/(2m_N)^2\right]^\frac{1}{2}}
         {1+\delta/m_N} 
 \, .
\label{eqn:kdelta}
\eeq
It is worth noting that $F_R$ is independent of energy by
definition.
Since the definition of $Z_\Delta$ is not unique \cite{gegelia},
interpreting $T_a\vec{S}\cdot\vec{k}\, F_R$ as the CM amplitude for
delta decay into $\pi N$ is questionable. We stress that $F_R$
is just another intermediate quantity that will be used later to
assemble the $\pi N$ scattering amplitude.

The LO and NLO of $F_R$ are the same as those of $F$:
\bea
F_R^{(0)} &=& F^{(0)} = \frac{h_{A,R}}{2 f_\pi}\; , \label{eqn:fR0}\\
F_R^{(1)} &=& F^{(1)} = 0 \; . \label{eqn:fR1}
\eea
At NNLO, the pion loops bring dependences on $\delta$ and
$k_\delta$. To the first approximation, $\delta = \omega(k_\delta) +
\mathcal{O}(k_\delta^2/m_N)$. In addition, $F_R^{(2)}$ receives
nontrivial one-loop corrections, which are evaluated on-shell by
definition, and contributions of the field renormalization constants,
\beq
F_R^{(2)} = \frac{1}{2}\left(Z_\pi^{(2)} + Z_N^{(2)} +Z_\Delta^{(2)}\right)
  F^{(0)} + F^{(2)}\left(\omega(k_\delta), \omega(k_\delta), k_\delta^2\right) 
= \frac{h_{A,R}}{2 f_\pi} \left(\varkappa + i \lambda \right)
\; , \label{eqn:fR2}
\eeq
where 
\beq
\varkappa \equiv \frac{k_\delta^2}{(4\pi f_\pi)^2} 
\left[d_{R} \frac{(4\pi f_\pi)^2}{h_{A,R}}
  \frac{m_\pi^2}{k_\delta^2}
+ \mathrm{Re} \mathcal{G}(m_\pi/k_\delta)
\right]\; , 
\label{eqn:ftilde}
\eeq
and
\beq
\lambda \equiv \frac{k_\delta^2}{(4\pi f_\pi)^2} 
          \mathrm{Im} \mathcal{G}(m_\pi/k_\delta) \; ,
\label{eqn:ftildetilde}
\eeq
with 
\bea
\mathcal{G}(x) &=& \frac{2}{3}\left(1+x^2\right)^{-\frac{1}{2}} 
\left\{
-\pi\left(g_A^2 - 
\frac{81}{16} {g_A^\Delta}^2\right)
x^3
+ 2\pi i \left(g_A^2 + \frac{1}{72}h_{A,R}^2\right) 
\right. \nonumber \\
&& \left.
+  
\left[
g_A^2 
  -\frac{1}{72}h_{A,R}^2 \left(13+15 x^2\right)+
\frac{81}{16} {g^\Delta_A}^2 \right]
\ln\left(\frac{\sqrt{1+x^2}-1}{\sqrt{1+x^2}+1}\right) 
\right\} \; 
\label{G(x)}.
\eea
There are two types of divergences arising from the one-loop corrections and
field renormalization constants: one is proportional to $\delta^2$ and the
other $m_\pi^2$. We can use the bare $h_A$ to absorb the $\delta^2$
divergence. In constructing $\mathcal{L}^{(2)}$
\eqref{eqn:lag2},  we already used the delta equation of motion
to turn the operator 
$-(N^\dagger\iv{T}\vec{S}\mathscr{D}_0^2{\Delta} +H.c.)
\bm{\cdot}\cdot\vec{\iv{D}}$ into
$\delta^2(N^\dagger\iv{T}\vec{S}{\Delta} +H.c.)$, 
which is subsequently absorbed into the $h_A$
operator. Therefore, it is appropriate to combine the $\delta^2$
divergence with the bare $h_A$, since there already is a $\delta^2$
piece in $h_A$.
A similar argument holds for the $d$ operator, which
is proportional to $m_\pi^2$ and can be renormalized by the 
$m_\pi^2$ divergence.
Equations~(\ref{eqn:fR0}), (\ref{eqn:fR2}), and
(\ref{eqn:ftilde}) should be viewed as the
definitions of the renormalized coupling constants $h_{A,R}$
and $d_{R}$. To make expressions compact, we will
drop the subscripts ``$_R$'' on  $h_{A,R}$
and $d_{R}$ in the rest of this paper.

Note that the logarithm in the function $\mathcal{G}(x)$
blows up as  $x\to 0$,
implying that an
infrared divergence would arise if one treated
$m_\pi/\delta$ as a higher-order effect, as in the power counting of 
Ref. \cite{pascalutsa}. Although this infrared divergence is not a
fundamental difficulty, it 
is convenient to avoid it by
considering $m_\pi \sim \delta$.

\subsection{$\Delta$ self-energy \label{sec:selfenergy}}

In this paper we employ the delta rest frame, where the on-shell 
delta energy is
\beq
E_\Delta(0) =\delta \; .
\label{EDrest}
\eeq
As we discussed, 
the delta self-energy produces a large effect
only inside the
resonance window, $|E - \delta| = \mathcal{O}(Q^3/\mqcd^2)$.
We will enforce this relation when
expanding $\Sigma_\Delta(E)$. The restriction on the energy domain
changes the order-index of $\Sigma_\Delta(E)$, since
the energy dependence of $\Sigma_\Delta(E)$ around $E = \delta$ is
always two powers smaller than $\Sigma_\Delta(\delta)$, \refeq{eqn:seexp}. 
We denote by
$\widehat{\Sigma}_\Delta^{(0)}, \widehat{\Sigma}_\Delta^{(1)}, \ldots$ 
the expansion of $\Sigma_\Delta(E)$ within the resonance window.

The diagrams contributing to $\Sigma_\Delta$ at LO, NLO, and NNLO can be found
respectively in \reffig{fig:se0}, \reffig{fig:se1}, and \reffig{fig:se2}. 
Detailed calculations of the delta
self-energy can be found in the literature,
for example Refs. \cite{jenkins, Banerjee:1994bk}.
Using Eqs.~\eqref{eqn:defgamma}, \eqref{eqn:deltarenorm}, \eqref{eqn:zd2}, 
and \eqref{eqn:dressedp}, 
the three lowest orders of
$\Sigma_\Delta(E)$ can then be written as
\bea
\widehat{\Sigma}_\Delta^{(0)}(\delta) 
&=& \frac{i}{2}\gamma^{(0)}(\delta) \; , \label{eqn:sigma0}\\
\widehat{\Sigma}_\Delta^{(1)}(\delta) 
&=& \frac{i}{2}\gamma^{(1)}(\delta) \; , \label{eqn:sigma1}\\
\widehat{\Sigma}_\Delta^{(2)}(E) 
&=& -\frac{Z_\Delta^{(2)}}{{S_\Delta^{(0)}(E)}} 
    +\frac{i}{2}\hat\gamma^{(2)}(E) \; ,
\label{eqn:sigma2}
\eea
where
\beq
\hat\gamma^{(2)}(E) \equiv 
\gamma^{(2)}(\delta)+ Z_\Delta^{(2)} \gamma^{(0)}(\delta)
+ (E - \delta){\gamma^{(0)}}'(\delta)\; .
\label{eqn:G2def}
\eeq
Due to the presence of $Z_\Delta^{(2)}$, $\widehat{\Sigma}_\Delta^{(2)}$ is
cutoff-dependent.
We should appreciate the fact that
the energy dependence of $\tr{Re}\Sigma_\Delta^{(0)}(E)$ is 
not present until NNLO, thanks to enforcing $|E - \delta| =
\mathcal{O}(Q^3/\mqcd^2)$. Had it shown up already in
$\widehat{\Sigma}_\Delta^{(0)}$, an otherwise redundant operator
$i \Delta^{\dagger}\mathscr{D}_0^3 \Delta$ would have been necessary
in order to absorb the divergence in $\Sigma_\Delta^{(0)}(E)$ that is
proportional to $E^3$.

The imaginary part of the delta self-energy, $\gamma(E)$, can be most
conveniently evaluated by cutting the intermediate states that
could be on-shell with an injected CM energy $E$
and then replacing those propagators with the Dirac
delta functions that enforce energy-momentum
conservation. This is of course equivalent to applying the optical theorem.

In LO and NLO delta self-energy
diagrams, the only potentially on-shell intermediate
state is a pion and a nucleon. 
The contribution of such an intermediate state to $\gamma(E)$ can
generally be written as
\bea
\gamma_{\pi N}(E) &=& Z_N Z_\pi \sum_{a} \int \frac{d^3l}{(2\pi)^3}
\frac{1}{2 \omega(l)}\,
{V_{\pi\,a}^{\text{CM}}}^{\dagger}(E)\, V_{\pi\,a}^{\text{CM}}(E) 
{}\times 2\pi \delta\left(
E - E_N(l) -\omega(l)\right)
\nonumber \\
&=& Z_N Z_\pi  \mathcal{N}(k) \frac{k^3}{6\pi}
\big{|} F\left(E, \omega(k), k^2 \right) \big{|}^2 \; ,
\label{eqn:def-gamma}
\eea
where, in the last line, $k$ satisfies
\beq
E=\omega(k) +E_N(k) \; ,
\label{eqn:EwEN}
\eeq
and the pre-factor is
\beq
\mathcal{N}(k) \equiv \frac{E_N(k) + m_N}{E +m_N} \; . \label{eqn:defN}
\eeq
For an on-shell delta,
use of Eq.~\eqref{eqn:defFR} yields 
\beq
\gamma_{\pi N}(\delta) = \mathcal{N}(k_\delta) \frac{k_\delta^3}{6\pi}
\frac{|F_R|^2}{Z_\Delta} \; .
\label{eqn:gammadelta}
\eeq

Substituting the LO expressions for $F$ (Eq.~\eqref{eqn:fR0}), 
$Z_\pi$ (Eq.~\eqref{Zpiexp0}) and $Z_N$ (Eq.~\eqref{ZNexp0}), 
\beq
\gamma^{(0)}(E)= \frac{h_A^2}{24 \pi f_\pi^2}  \mathcal{N}(k) k^3  \; .
\label{eqn:gamma0E}
\eeq
Using in
$\mathcal{N}(k_\delta)\,k_\delta^3$ the exact kinematic relation between 
$k_\delta$ and $\delta$, Eq.~\eqref{eqn:kdelta},
\beq
\gamma^{(0)}(\delta) =
\frac{h_A^2}{24 \pi f_\pi^2} \left(\delta^2 -m_\pi^2\right)^{\frac{3}{2}}
\left[1+ \delta/m_N + (\delta^2 -m_\pi^2)/(2m_N)^2\right]^\frac{3}{2} 
\frac{1+ \delta/m_N + (\delta^2 -m_\pi^2)/2m_N^2}{\left(1+\delta/m_N\right)^5}
\; ,
\label{eqn:gamma0}
\eeq
a relation known from isobar models \cite{ericson}. 
Analogously, from the NLO expressions for $F$ (Eq.~\eqref{eqn:fR1}),
$Z_\pi$ (Eq.~\eqref{Zpiexp1}), and $Z_N$ (Eq.~\eqref{ZNexp1}),
\beq
\gamma^{(1)}(\delta)=0 \; .
\label{eqn:gamma01}
\eeq

The strict heavy-baryon expansion of $\mathcal{N}(k_\delta)\,k_\delta^3$ gives
\beq
\gamma^{(0)}(\delta)+\gamma^{(1)}(\delta)+   \cdots
= \frac{h_A^2}{24 \pi f_\pi^2} \left(\delta^2 -m_\pi^2\right)^{\frac{3}{2}}
\left[1 -\frac{5\delta}{2 m_N} + \frac{42\delta^2 - 7m_\pi^2}{8 m_N^2} -
  \frac{7\delta\left(22\delta^2 - 7m_\pi^2\right)}{16m_N^3} +
  \cdots
\right]
\; .
\label{eqn:hbegamma0}
\eeq
The first term is the well-known heavy-baryon limit \cite{jenkins}.
With $\delta \sim 300$ MeV, the first, Galilean
correction is $\sim -80\%$ due to the relatively large numerical factor 
in front of $\delta/m_N$. This expansion, nonetheless, still converges as long
as $\delta < m_N$. More importantly, the slowness of this expansion is
not reflected in the EFT expansion of the amplitude. As we will show
later, 
to NLO the $\pi N$ amplitude only depends on $\delta$
and $\gamma^{(0)}+\gamma^{(1)}$, 
{\it i.e.}, it is a simple Breit-Wigner formula. 
Enforcing the heavy-baryon expansion in 
$\mathcal{N}(k_\delta)\,k_\delta^3$ does not result
in an amplitude with a different functional form, but it does lead to a
different value of $h_A$. 
This only means that the value of $h_A$ 
depends, because of the slow convergence, on how
$Q/m_N$ corrections are treated.
In addition, there are
higher-order corrections to $\gamma$ contributed by the NNLO $\pi N
\Delta$ vertex function, which includes one-loop corrections 
and an undetermined LEC 
($d$).
Since there is no \textit{a priori} evidence that
these undetermined corrections are 
smaller than those that are
proportional to $1/m_N^3$ and higher, 
keeping $Q/m_N$ terms 
to all orders does not have any deep significance.

In the following, we consider two cases.
In one case, having explained the validity of the heavy-baryon formalism
despite the large factor of $-5\delta/2m_N$, 
we carry out the expansion of $\mathcal{N}(k) k^3$.
We refer to this as the strict heavy-baryon expansion.
In the other case, 
we do not 
expand $\mathcal{N}(k) k^3$, 
because
\textit{(i)}
it is a convenient way to include the required terms up to $1/m_N^2$;
\textit{(ii)}
it allows a meaningful comparison of LECs with the literature,
where a similar resummation is performed \cite{ellis-tang, pascalutsa};
and 
\textit{(iii)} whenever desirable,
a strict heavy-baryon expansion in $Q/m_N$ can easily be worked out.
We refer to this second case as the semi-resummation.

When evaluating $\gamma^{(2)}(\delta)$, there are a few NNLO delta
self-energy diagrams in which an intermediate state of two pions
and a nucleon ($\pi \pi N$) could be on-shell if being cut on the
middle-nucleon internal line. These diagrams are labeled 
(a) to (h) in \reffig{fig:se2}.
To estimate the contribution of $\pi \pi N$ to
$\gamma^{(2)}(\delta)$, we first notice that the phase space for such
an intermediate state to go on-shell is so small that even pions are
nonrelativistic, having three-momenta 
$\tilde{Q} \sim \sqrt{m_\pi(\delta -2m_\pi)} \sim 40$ MeV $\ll m_\pi$. 
We can use this fact
to refine once again the standard ChPT power counting. To be
definite, let us look at the two-loop diagram labeled (a) in
\reffig{fig:se2}. In contrast with the generic situation, we
should now replace $\tilde{Q}$ for $Q$ where pion three-momenta
appear in the loops (vertices, propagators, integrals).
Also, the energy of the
nucleon is of $\mathcal{O}(\tilde{Q}^2/m_N)$ rather than $Q$. Overall,
these changes in the power counting bring a suppression of roughly
order $(\tilde{Q}/m_\pi)^{\sim 7} \sim 10^{-3} - 10^{-4}$.
This somewhat crude
estimate is justified by a phase-shift analysis \cite{gwpwa}. The
unitarity of the $S$ matrix suggests that
the opening of the $\Delta \to \pi \pi N$
channel brings ``inelasticities'' in $\pi N$ scattering.
However, Ref.~\cite{gwpwa} gives
inelasticities only of order of magnitude $10^{-3}$ in the delta region.
Therefore, though they are formally NNLO, the $\pi \pi N$ 
contributions are suppressed by the 
``accidental'' closeness of $\delta$ to $2m_\pi$.
Numerically we can thus safely neglect $\pi \pi N$
contributions, 
\beq
\gamma_{\pi\pi N}(\delta) \simeq 0 \; ,
\label{eqn:gammapipidelta}
\eeq
and consider only $\pi N$ contributions.

With this approximation and 
\refeq{eqn:fR2},
$\gamma^{(2)}(\delta)$ is given by
\beq
\gamma^{(2)}(\delta) =
\left(2\varkappa-Z_\Delta^{(2)}\right) \gamma^{(0)}(\delta)\; ,
\label{eqn:gamma2}
\eeq
where $\varkappa$ is given in Eq.~\eqref{eqn:ftilde}.
Inserting \refeq{eqn:gamma2} into \refeq{eqn:G2def}
one finds
\beq
\hat\gamma^{(2)}(E) = 2\varkappa \gamma^{(0)}(\delta) +
{\gamma^{(0)}}'(\delta) (E - \delta)
\; ,
\label{eqn:G2}
\eeq
completing the calculation of the delta self-energy to NNLO.

\section{$\pi N$ scattering amplitude\label{sec:amps}}

In this section we put together the various 
one-$\Delta$-irreducible ingredients calculated in the previous section.
We first review the kinematics in $\pi N$ scattering,
before constructing the $T$ matrix in the various channels.

\subsection{Kinematics\label{sec:kin}}
In the CM frame, we denote the initial (final) pion momentum by
$\vec{k}$ (${\vec{k}}'$), the initial (final) pion isospin index by
$a$ ($a'$), and the initial (final) nucleon spin $z$-component and
isospin third-component by
$\sigma$ ($\sigma'$) and $\tau$ ($\tau'$), respectively. 
The CM energy, denoted by $E$,
is given in terms of
the pion (nucleon) energy $\omega(k)$ ($E_N(k)$)
by \refeq{eqn:EwEN}, with $k \equiv |\vec{k}|$.

The $T$ matrix is related to the $S$ matrix by
\beq
S = 1 + i T \; .
\label{eqn:snt}
\eeq
Asymptotic pion states are normalized so that
\beq
\langle \iv{\pi}, {\vec{k}}'\, a' | \iv{\pi}, \vec{k}\, a \rangle = 
2 \omega(|\vec{k}|)\, (2\pi)^3\,
\delta^{(3)}(\vec{k}-{\vec{k}}')\, \delta_{{a'}a}\; ,
\label{eqn:pi-nor}
\eeq
while for nucleon states,
\beq
\langle N, {\vec{k}}'\, \sigma'\, \tau' | N, \vec{k}\, \sigma\, \tau
\rangle =  (2\pi)^3\, \delta^{(3)}(\vec{k}-{\vec{k}}')\,
\delta_{\sigma' \sigma} \delta_{\tau' \tau} \; .
\label{eqn:nuc-nor}
\eeq
The scattering amplitudes, $\mathcal{A}_{a' a}$, are the elements of the
$T$ matrix between the asymptotic pion and nucleon states and can be
written as $2 \times 2$
matrices in nucleon spin and isospin indices,
\beq
\left(\mathcal{A}_{a'a}\right)_{\sigma' \sigma\, , \tau' \tau}
     (\vec{k}', \vec{k}) \equiv 
\langle {\vec{k}}'\, a'\, \sigma'\, \tau' | T | \vec{k}\, a\,
\sigma\, \tau \rangle \; .
\eeq

We normalize
the spin-orbital projector $\mathbb{P}_{jl}$ for a total angular
momentum $j$ and an orbital angular momentum $l$ so that
\beq
\int d\Omega_{\hat{k}''}
\mathbb{P}_{j'l'}(\hat{k}',\hat{k}'')
\mathbb{P}_{jl}(\hat{k}'',\hat{k}) = \delta_{j'j} \delta_{l'l}
\mathbb{P}_{jl}(\hat{k}',\hat{k}) \; ,
\eeq
with $d\Omega_{\hat{k}''}$ the area element on a unit three-sphere
spanned by $\hat{k}''$. The $\mathbb{P}_{jl}$s
with lowest $j$s and $l$s are
\bea
\mathbb{P}_{\frac{1}{2} 0}(\hat{k}',\hat{k}) &=& \frac{1}{4\pi}\; ,\\
\mathbb{P}_{\frac{1}{2} 1}(\hat{k}',\hat{k}) &=&
\frac{1}{4\pi} \left[ \hat{k}'\cdot\hat{k} +
  i\left(\hat{k}'\times\hat{k}\right)\cdot\vec{\sigma} \right]\; ,\\
\mathbb{P}_{\frac{3}{2} 1}(\hat{k}',\hat{k}) &=&
\frac{1}{4\pi} \left[ 2\hat{k}'\cdot\hat{k} -
  i\left(\hat{k}'\times\hat{k}\right)\cdot\vec{\sigma} \right]
\; .
\eea
The isospin projector $\mathbb{I}_t$ for a total isospin $t$ is
normalized so that
\beq
\sum_c \mathbb{I}_{t'}(a', c) \mathbb{I}_{t}(c, a) = \delta_{t' t}
\mathbb{I}_t(a', a) \; .
\eeq
There are only two different $\mathbb{I}_t$s in $\pi N$ scattering:
\bea
\mathbb{I}_{\frac{1}{2}}(a', a) &=& \frac{1}{3} \left(\delta_{a'a} +
  i\epsilon_{a'ac}\tau_c \right)\; , \\
\mathbb{I}_{\frac{3}{2}}(a', a) &=& \frac{1}{3} \left(2\delta_{a'a} -
  i\epsilon_{a'ac}\tau_c \right)\; .
\eea

The normalization factor between the angular-momentum
eigenstates and the asymptotic states defined in
Eqs.~(\ref{eqn:pi-nor}) and (\ref{eqn:nuc-nor}) can be found in many
textbooks (see, \textit{e.g.}, Ref.~\cite{weinbergbook}). Without
showing the tedious details, we simply state that $\mathcal{A}_{a' a}$
is related to the phase shifts, $\theta_{jlt}(E)$, as follows:
\beq
i \mathcal{A}_{a'a}(\vec{k}', \vec{k}) \equiv  \frac{8\pi^2}{k\,
  \mathcal{N}(k)} \sum_{j\,l\,t}
\mathbb{P}_{jl}(\hat{k}', \hat{k}) \mathbb{I}_t(a', a)
\left\{\exp\left[2i\theta_{jlt}(E)\right] - 1\right\} \; ,
\label{eqn:AT} 
\eeq
where $\mathcal{N}(k)$ is defined in \refeq{eqn:defN}.
The partial-wave $T$-matrix elements are expressed in terms of
$\theta_{jlt}(E)$ as
\beq
T_{jlt}(E) \equiv -i \left\{\exp\left[ 2i\theta_{jlt}(E) \right] - 1\right\} 
\; .
\label{eqn:ps-def}
\eeq

In the following we will use
a more conventional notation for a specific partial wave: $l_{2t, 2j}$. For
example, $P_{13}$ refers to the $l=1$ ($P$ wave), $t=1/2$, and $j=3/2$.

\subsection{$T$ matrix\label{sec:t}}
Now we 
collect all the pieces from Sec.~\ref{sec:pre}
to build the $\pi N$ scattering
amplitude in the various waves, Eq.~\eqref{eqn:ps-def}. 
Here the exact relation between 
$E$ and $k$ (\refeq{eqn:EwEN}) is assumed,
meaning that certain trivial, kinematic $k/m_N$ 
terms are resummed ---what we refer to as semi-resummation. 
In the next section, Sec.\ref{sec:tstrict},
we specialize to the strict heavy-baryon expansion.

At LO ($Q^{-1}$) there is only a pole diagram,
\reffig{fig:powct}(A), which contributes only to the $P_{33}$ wave.
{}From Eqs.~\eqref{eqn:dressedp} and \eqref{eqn:sigma0},
\beq
T_{P_{33}}^{\text{LO}} = -\gamma^{(0)}(\delta)\,S_\Delta^{(0)}\,
=-\frac{\gamma^{(0)}(\delta)}{E - \delta +
  i\gamma^{(0)}(\delta)/2} 
\left[ 1 +\mathcal O\left(\frac{Q}{\mqcd}\right) \right] \; ,
\label{eqn:TP33LO}
\eeq
where
$\gamma^{(0)}(\delta)$ is given by Eq.~\eqref{eqn:gamma0}. 
This is of the form \eqref{eqn:bwback} with 
a resonance at $E_R=\delta$,
an energy-independent width $\Gamma= \gamma^{(0)}(\delta)$,
and no background, $T_B=0$.
The two independent parameters can be taken to be $\delta$ and 
$h_A$. 
This is same result as in any isobar model with the simplest
contribution to the width resummed \cite{ericson}.

In next order ($Q^{0}$), there appear $Q/m_N$ corrections
to the pole diagram, \reffig{fig:powct}(B),
which contribute via corrections to the delta self-energy, \refeq{eqn:sigma1}.
The NLO amplitude has the same form as LO,
\beq
T_{P_{33}}^{\text{NLO}} = 
-\frac{\gamma^{(0)}(\delta)+\gamma^{(1)}(\delta)}{E - \delta +
  i\left[\gamma^{(0)}(\delta)+\gamma^{(1)}(\delta)\right]/2} 
\left[ 1 +\mathcal O\left(\frac{Q^2}{\mqcd^2}\right) \right] \; ,
\label{eqn:TP33NLO}
\eeq
where
$\gamma^{(1)}(\delta)$ vanishes, as given by Eq.~\eqref{eqn:gamma01},
when we do not expand
kinematic relations in powers of $\delta/m_N$.

The NNLO ($Q^1$) corrections are more complicated. 
The cutoff dependences in $V_\pi^{(2)}$,
$\Sigma_\Delta^{(2)}$, $Z_N^{(2)}$, and $Z_\pi^{(2)}$ make several
NNLO pole diagrams (labeled from (1) to (4) in \reffig{fig:powct}(C))
divergent. 
The remaining diagrams in \reffig{fig:powct}(C) involve only
$Q/m_N$ corrections.
To see that the divergences of these diagrams in fact
cancel each other, 
consider the sum of these pole diagrams,
\bea
T^{\tr{pole}(2)}_{P_{33}} &=& 
-\gamma^{(0)}(\delta)\,S_\Delta^{(0)}\,
 \left[\frac{4f_\pi}{h_A} 
       F^{(2)}\left(\omega(k_\delta),\omega(k_\delta),k_\delta^2\right) 
     - S_\Delta^{(0)}\,\widehat\Sigma_\Delta^{(2)}(E) 
     + \frac{\gamma^{(0)'}(\delta) (E-\delta)}{\gamma^{(0)}(\delta)}
\right.\nonumber\\
&& \left. \qquad\qquad \qquad   
     + Z_N^{(2)} + Z_\pi^{(2)} \right]\; .
\eea
Using Eqs.~(\ref{eqn:fR2}), (\ref{eqn:sigma2}), and (\ref{eqn:G2}),
and defining 
\beq
T_B(\delta) \equiv 2\lambda \; , 
\label{eqn:Bdelta}
\eeq
we find 
\beq
T^{\tr{pole}(2)}_{P_{33}} = 
-\gamma^{(0)}(\delta)\,S_\Delta^{(0)}\,
\left[- \frac{i}{2}S_\Delta^{(0)}\, \hat\gamma^{(2)}(E) +
\frac{\hat\gamma^{(2)}(E)}{\gamma^{(0)}(\delta)} + i T_B(\delta) \right] \; ,
\label{eqn:pole2}
\eeq
where no cutoff dependence is present.

Now summing up all the pole diagrams up to order $Q^1$, 
\bea
T^{\tr{pole}}_{P_{33}} &=& T_{P_{33}}^{\text{LO}}+ T^{\tr{pole}(2)}_{P_{33}} 
\nonumber\\
&=&-\gamma^{(0)}(\delta)\,S_\Delta^{(0)}\,
\left[1 - \frac{i}{2}S_\Delta^{(0)}\, \hat\gamma^{(2)}(E) +
\frac{\hat\gamma^{(2)}(E)}{\gamma^{(0)}(\delta)} + i T_B(\delta) \right] 
\left[1+ \mathcal O\left(\frac{Q^3}{\mqcd^3}\right) \right] \; .
\eea
Within the stated error, we can 
resum the corrections,
\beq
T^{\tr{pole}}_{P_{33}} = 
-\frac{\gamma^{(0)}(\delta) + \hat\gamma^{(2)}(E)}{E - \delta +
i \left[\gamma^{(0)}(\delta) + \hat\gamma^{(2)}(E)\right]/2} 
\left[1 + iT_B(\delta)\right] 
\left[1 +\mathcal{O}\left(\frac{Q^3}{\mqcd^3}\right) \right]
\; .
\label{eqn:polep33}
\eeq

The result so far relied on an expansion around the resonance,
to be joined with a description of the off-pole region.
As we discussed in Sec.~\ref{sec:sewing}, \reffig{fig:powct}(D) 
can be accounted for off-pole by allowing the ``full'' energy dependence in the
external legs in \reffig{fig:powct}(A). This eventually amounts to
replacing in \refeq{eqn:polep33} 
$\gamma^{(0)}(\delta)$ with $\gamma^{(0)}(E)$ and, in
addition, $\hat\gamma^{(2)}(E)$ with 
$\hat\gamma^{(2)}(\delta)=2 \varkappa \gamma^{(0)}(\delta)$ in order to
avoid over-counting the energy dependence.
Moreover, to the order we are working, we can instead
replace $\hat\gamma^{(2)}(\delta)$ with 
$[(1+\varkappa )^2 -1] \gamma^{(0)}(E)$.
We then arrive at
\beq
T^{\text{pole/off-pole}}_{P_{33}} = -\frac{\Gamma(E)}{E
  - \delta + i\Gamma(E)/2} \left[1 +i T_B(\delta)\right] 
\left[1 +\mathcal{O}\left(\frac{Q^3}{\mqcd^3}\right) \right]
\; ,
\label{poleoffpole}
\eeq
where, using Eq.~\eqref{eqn:gamma0E},
\beq
\Gamma(E) =\frac{\left[h_A(1+\varkappa)\right]^2}{24 \pi f_\pi^2} 
k^3\mathcal{N}(k)
\; . 
\label{eqn:Gamma(E)}
\eeq
Equation \eqref{poleoffpole} resembles \refeq{eqn:bwback} except for the
absence of an additional background term $T_B(\delta)$,
which would sit outside the pole term.
The one-$\Delta$-irreducible trees at order $Q^1$ will provide the
remaining piece expected from unitarity, as we will now show.

In fact, in order to complete order $Q^1$, we
need to include the remaining trees, \reffig{fig:powct}(E), which
are one-$\Delta$-irreducible. 
Their contributions to the $P_{33}$ channel are found to be
\beq
T^{\tr{tree, 1$\Delta$I}}_{P_{33}} = T_B(E) \; ,
\label{eq:treeP33}
\eeq
with
\beq
T_B(E) = \frac{k^3 \mathcal{N}(k)}{6\pi f_\pi^2}\left(\frac{g_A^2}{E} +
  \frac{1}{36}\frac{h_A^2}{E+\delta} \right) 
\; .
\label{eq:B(E)}
\eeq
This reduces in leading-order in $Q/m_N$ 
to $T_B(\delta)$ defined in \refeq{eqn:Bdelta} 
when $E= \delta$,
which is exactly the missing piece expected from
unitarity. This is certainly not a miracle as EFT is supposed to
reproduce the unitarity and analyticity of the $S$ matrix order by order. 
However, it
does confirm the consistency of our power counting from a
particular perspective.

Now one can sum up pole \eqref{poleoffpole}
and tree \eqref{eq:treeP33}
contributions to obtain the $P_{33}$
partial-wave amplitude. Since the difference between $E$ and $\delta$
is higher order in the NNLO pole term \eqref{eqn:pole2},
\beq
T_{P_{33}}^{\text{NNLO}} = 
 -\frac{\Gamma(E)}{E - \delta + i\Gamma(E)/2} \left[1 +i T_B(E)\right] 
+ T_B(E) + \mathcal O\left(T_{P_{33}}^\text{LO}\,\frac{Q^3}{\mqcd^3}\right)
\; . 
\label{eqn:Tp33}
\eeq
Again we recover a Breit-Wigner form \eqref{eqn:bwback},
but now 
with an energy-dependent width, \refeq{eqn:Gamma(E)}, 
and an energy-dependent background, \refeq{eq:B(E)}.
The NNLO amplitude involves 
four independent parameters, $\delta$, $h_A$, $g_A$, and $\varkappa$.
The EFT thus provides specific energy dependences for
$\Gamma$ and $T_B$ through
the two extra parameters that appear at NNLO, $g_A$ and $\varkappa$.
Our amplitude is similar to the one from isobar models \cite{ericson},
except that $h_A$ in $\Gamma(E)$ gets corrected by a factor of $1+\varkappa$.

Equation \eqref{eqn:Tp33} reduces to 
\refeq{eqn:bwback} with energy-{\it in}dependent
$\Gamma(\delta)$ and $T_B(\delta)$ when $|E - \delta|$ is small enough.
For that to happen, the
contribution from $T_B(\delta)$ must overpower that of the energy dependence 
of $\Gamma(E)$, which according to \refeq{eqn:pole2} requires
the size of the energy window to be smaller than the half width.
In terms of the scales in the problem,
we can define a ``Breit-Wigner window'' 
$|E -\delta| \sim \mathcal{O}(Q^4/\mqcd^3)$
where \refeq{eqn:bwback} holds approximately.
Outside this small window, the more general form \eqref{eqn:Tp33}
should be used.

Other channels are easy to calculate from the
one-$\Delta$-irreducible tree diagrams in \reffig{fig:powct}(E).
For the remaining $P$-wave channels,
\beq
T_{P_{13}}^{\text{NNLO}} = T_{P_{31}}^\text{NNLO} 
=\frac{1}{4} T_{P_{11}}^\text{NNLO}=
-\frac{k^3 \mathcal{N}(k)}{12 \pi f_\pi^2}
\left(\frac{g_A^2}{E} - \frac{2}{9}\frac{h_A^2}{E+\delta}\right)
\left[1 + \mathcal O\left(\frac{Q}{\mqcd}\right) \right]
\; . \label{eqn:Tp131} 
\eeq

The one-$\Delta$-irreducible contributions, 
Eqs. \eqref{eq:B(E)} and \eqref{eqn:Tp131}, are, of course, related to 
amplitudes found
in the literature.
They are part of the simplest tree-level delta-isobar model 
\cite{ericson}, where 
the relations among the NNLO amplitudes in $P_{13}$, $P_{31}$, and
$P_{11}$ were also noticed.
These contributions are LO in both deltaless
and deltaful EFTs when one uses the standard power counting.
Ignoring the delta contribution, namely
the terms proportional to $h_A^2$,
one reproduces
the deltaless EFT results in Refs.~\cite{mojzis, fettes-deltaless-is}.
Including the delta contribution, our results agree
with those in the deltaful EFT given in 
Refs.~\cite{threshold-delta1,threshold-delta2,vijay}.

The seagull diagram in \reffig{fig:powct}(E)
(the Weinberg-Tomozawa term) contributes to the $S_{11}$
and $S_{31}$ channels. Since the delta does not contribute to these waves,
our results reduce to the well-known ChPT expressions at standard LO
\cite{weinbergbook,BerKaiMei}.
In the following, we focus on the $P$-waves.

\subsection{Strict heavy-baryon expansion\label{sec:tstrict}}

We have partially resummed $Q/m_N$ terms in the 
amplitudes shown above.
More precisely, 
this semi-resummation consists of keeping the
exact kinematic relations 
\eqref{eqn:kdelta} between $\delta$ and $k_\delta$ in \refeq{eqn:gammadelta},
and \eqref{eqn:EwEN} between $k$ and $E$ 
in Eqs.~\eqref{eqn:Gamma(E)}, \eqref{eq:B(E)}, and \eqref{eqn:Tp131}. 
If, instead, one enforces a strict $1/m_N$ expansion,
there are some small changes in the results.

In the $P_{33}$ channel, at LO and NLO the amplitude is still given
by Eqs.~\eqref{eqn:TP33LO} and \eqref{eqn:TP33NLO}, except that
$\gamma^{(0)}(\delta)$ and $\gamma^{(1)}(\delta)$ are
given by the first two terms in \refeq{eqn:hbegamma0}.
At NNLO, 
again the amplitude remains in the same form as \refeq{eqn:Tp33} but with
different expressions for the width,
\beq
\Gamma(E) = \frac{\left[h_A(1+\varkappa)\right]^2}{24 \pi f_\pi^2} 
\left(E^2 -m_\pi^2\right)^{\frac{3}{2}}
\left[1-\frac{5 E}{2 m_N} + \frac{42E^2 - 7m_\pi^2}{8 m_N^2}
\right]\;  , 
\label{eqn:HBGamma(E)}
\eeq
and for the background,
\beq
T_B(E) = \frac{\left(E^2 -  m_\pi^2\right)^{\frac{3}{2}}}{6\pi f_\pi^2}
\left(\frac{g_A^2}{E} + \frac{1}{36}\frac{h_A^2}{E+\delta} \right) \; .
\eeq
Like for $T_B(E)$, in other $P$-waves we simply replace 
$k$ with $\sqrt{E^2 - m_\pi^2}$:
\beq
T_{P_{13}}^{\text{NNLO}} = T_{P_{31}}^\text{NNLO} =\frac{1}{4}
T_{P_{11}}^\text{NNLO} = -\frac{(E^2-m_\pi^2)^{\frac{3}{2}}}{12 \pi f_\pi^2} 
           \left( \frac{g_A^2}{E}-\frac{2}{9}\frac{h_A^2}{E+\delta}\right) 
\left[1 + \mathcal O\left(\frac{Q}{\mqcd}\right) \right]
\; .
\eeq

The difference between these expressions and the corresponding ones in the 
previous section is of higher order. Comparing the effects of the two 
sets of formulas
gives a further estimate of the size of higher-order corrections.

\subsection{Scattering volumes \label{sec:scattvols}}

Although we do not aim at a precise description of the threshold region,
we can extract from our calculation the $S$-wave scattering lengths
and the $P$-wave scattering volumes, which are related to the 
amplitudes at zero energy. 

Up to order $Q^1$ in the off-pole region,
the scattering lengths are, of course, identical to the venerable
current-algebra results \cite{originalS,weinbergbook}.
The scattering volumes, with the semi-resummation, are
\bea
a_{P_{33}} = \frac{g_A^2}{12 \pi f_\pi^2 m_\pi} 
  \left[1 +\left(\frac{\sqrt{2}h_A}{3g_A}\right)^2
  \frac{m_\pi/\delta}{1- m_\pi^2/\delta^2}
  \left(\frac{5}{4} + m_\pi/\delta\right)\right]\left(1+m_\pi/m_N\right)^{-1}
\; , 
\label{eqn:ap33} \\ 
a_{P_{13}} = a_{P_{31}} = \frac{1}{4} a_{P_{11}} =
-\frac{g_A^2}{24 \pi f_\pi^2 m_\pi}
\left[1-\left(\frac{\sqrt{2}h_A}{3g_A}\right)^2
\frac{m_\pi/\delta}{1+m_\pi/\delta}\right]\left(1+m_\pi/m_N\right)^{-1}
\; .
\label{eqn:apnd}
\eea
In the strict expansion, the factor of $(1+m_\pi/m_N)^{-1}$ should
be dropped.
The extra suppression $m_\pi/\delta$ of the delta contributions at threshold is
evident, and when $m_\pi/\delta\to 0$ the delta decouples, as it should. 
In the opposite limit, $m_\pi/\delta\to \infty$, the delta contributions
grow and the scattering volumes vanish (to order $Q^1$) 
if $h_A/g_A=3/\sqrt{2}$,
as is the case in the large-$N_c$ limit \cite{largenc}.
(Of course our results do not direct apply to this limit,
where the delta and other degrees of freedom are degenerate with the nucleon
and do not decay into a nucleon and a pion.)
The real world is in-between these two limits:
since $m_\pi/\delta\sim 1/2$ and $h_A/g_A\sim 3/\sqrt{2}$ 
(as we are going to see),
delta's contributions are neither terribly small nor completely
opposite to the nucleon's.

\section{$P$-wave Phase Shifts\label{sec:psa}}

To test our EFT result, we compare it with the phase-shift
analysis (PSA) by the George Washington (GW) group~\cite{gwpwa},
which bridges over the delta resonance.
With parameters extracted from the fit, we compare the
resulting values of the scattering volumes with those
obtained by a PSA that focuses on lower energies \cite{matsinos}.

We first extract the phase shifts out of the EFT amplitude
in such a 
way that unitarity is preserved perturbatively.
For that, we use \refeq{eqn:ps-def}, expanding both the phase
shifts and the $T$ matrix in powers of $Q/\mqcd$:
denoting by a superscript $^{(n)}$ the corresponding power,
\beq
\exp\left[ 2i \sum_n \theta^{(n)} \right] = 1 +  i \sum_n
  T^{(n)} \; .
\eeq
Specifically, in the $P_{33}$ channel at LO,
\beq
\theta_{P_{33}}^{\text{LO}} =
\tr{arccot}\left[\, 2\frac{\delta -E}{\gamma^{(0)}(\delta)}\right]
\; ,
\label{eqn:thtp33LO}
\eeq
at NLO,
\beq
\theta_{P_{33}}^{\text{NLO}} =
\tr{arccot}\left[\, 
2\frac{\delta -E}{\gamma^{(0)}(\delta)+\gamma^{(1)}(\delta)}\right]
\; ,
\label{eqn:thtp33NLO}
\eeq
and at NNLO,
\beq
\theta_{P_{33}}^{\text{NNLO}} =
\tr{arccot}\left[\, 2 \frac{\delta -E}{\Gamma(E)} \right]
+ \frac{T_B(E)}{2}
\; .
\label{eqn:thtp33NNLO}
\eeq
In the other $P$ channels,
\beq
\theta_{P_{13}}^{\text{NNLO}} = \theta_{P_{31}}^{\text{NNLO}}
=\frac{1}{4} \theta_{P_{11}}^{\text{NNLO}}
=\frac{1}{2} T_{P_{13}}^{\text{NNLO}} = \frac{1}{2} T_{P_{31}}^\text{NNLO} 
=\frac{1}{8} T_{P_{11}}^\text{NNLO} \; .
\label{eqn:thtp131}
\eeq

Several LECs enter these EFT results.
A number of them can be determined from other processes,
such as pion decay.
We adopt the following values: 
$m_\pi=139$ MeV, $m_N=939$ MeV, 
$g_A = 1.29$, and $f_\pi=92.4$ MeV. 
The value for $g_A$ is obtained using the Goldberger-Treiman relation, 
$g_A=4 \sqrt{\pi} f_\pi f /m_{\pi^+}$,
from the pion-nucleon
coupling constant $f$ determined very precisely 
in the Nijmegen PSA of two-nucleon
data \cite{BoM}.
Note that this includes
the chiral-symmetry-breaking corrections to
the pion-nucleon vertex (the Goldberger-Treiman discrepancy),
which 
only appear two orders  
higher than the highest order we are working at.
We have verified that our results are not very sensitive to such small 
corrections, and 
that using the chiral-limit value does not affect any of our conclusions.
In principle we could determine $|g_A/f_\pi|$ from
our fit, but that would distract from our main objective,
the description of the delta resonance.
The LECs germane to delta-resonance physics are $\delta$, $h_A/f_\pi$, and 
$\varkappa$. These LECs 
are to be determined from low-energy reactions involving the
delta, and we do so here for $\pi N$ scattering.

Our strategy of fitting is to determine the free parameters, $\delta$,
$h_A$, and $\varkappa$ from the $P_{33}$ phase shifts around the delta
peak and then predict the phase shifts at lower energies in all
$P$ waves. 
As it is clear from Eqs.~\eqref{eqn:thtp33LO} and 
\eqref{eqn:thtp33NLO}, the LO and NLO $P_{33}$ phase shifts have
the same functional dependence on $E$:
LO and NLO only differ by Galilean corrections buried in the 
width.
In the following figures we will display  Eqs.~\eqref{eqn:thtp33LO} and
\eqref{eqn:thtp33NNLO} by lines.
We fit our curves to the results of
the GW PSA \cite{gwpwa}, which we indicate by dots.
The two (four) points around $\theta_{P_{33}}=\pi/2$ used to determine
the two (three) parameters at LO (NNLO) are explicitly marked.

The difference between Eqs.~\eqref{eqn:thtp33LO} and \eqref{eqn:thtp33NNLO}
is itself an estimate of the systematic
theoretical error at LO.
We estimate the 
error at NNLO 
by power counting the higher-order corrections that
are neglected. 
As shown in \refeq{eqn:Tp33}, the
systematic error in the $P_{33}$ channel is of size 
$\mathcal{O}(T_{P_{33}}^\text{LO}\, Q^3/\mqcd^3)$,
while in the other channels it is of
$\mathcal{O}(T_{\text{Non}-\Delta}^\text{NNLO}\, Q/\mqcd)$,
see \refeq{eqn:Tp131}. 
To make the estimate more concrete,
we take for $\mqcd$ the scale associated with 
the lowest-lying baryon integrated out in the EFT,
the Roper resonance of mass $m_{R} = 1440$ MeV.
For simplicity we take the factor to
be $m_R-m_N$ in all channels;
while this could be a conservative estimate of the error since
Roper contributions are likely to be small in channels other than
$P_{11}$, it might still be representative
in the $P_{33}$ channel, where the next resonance, the $\Delta (1600)$,
lies above the delta a similar distance.
Since
$P$ waves other than $P_{33}$ are predictions
that apply throughout the range of energies we consider,
$Q$ should be given by the pion momentum, so the errors in
these channels can be generically estimated as
\beq
\Delta T_{\text{Non}-\Delta} = \pm T^{\text{NNLO}}_{\text{Non}-\Delta}
\frac{k}{m_R - m_N} \; .
\eeq
In the $P_{33}$ channel, on the other hand,
we take $Q$ as a measure of the deviation from the resonance,  
because we would like to make the estimated uncertainty
vanish at $E = \delta$ where we fit to the PSA inputs. 
We first estimate the errors of $\Gamma(E)$ and $T_B(E)$, then
calculate the systematic error of the 
amplitude using \refeq{eqn:Tp33}. In doing so the unitarity condition is 
explicitly preserved. 
The errors of $\Gamma(E)$ and $T_B(E)$ 
should be proportional to $E - \delta$.
Near the resonance, where $\Gamma(E)$ matters, $E - \delta \sim Q^3/\mqcd^2$.
The third-order correction to $\Gamma(E)$ is of
$\mathcal{O}(\gamma^{(0)}(\delta)\, Q^3/\mqcd^3)$, which then could be
estimated by
\beq
\Delta \Gamma(E) =\pm 
\Gamma(E) \frac{E - \delta}{m_{R} - m_N} \; .
\eeq
The higher-order corrections to $T_B(E)$ come from
one-$\Delta$-irreducible tree diagrams with one $\nu = 1$ vertex,
and hence are of $\mathcal{O}(T_B(E)\, Q/\mqcd)$.
Since, like in other channels, 
these one-$\Delta$-irreducible diagrams apply throughout the
low-energy region, $E - \delta \sim Q$ and
\beq
\Delta T_B(E) =\pm T_B(E) \frac{E - \delta}{m_R - m_N} \; .
\eeq
The systematic-error bands of the NNLO results 
are shown in the following figures
as shaded regions.

\begin{figure}
\centering
\includegraphics[scale=0.75]{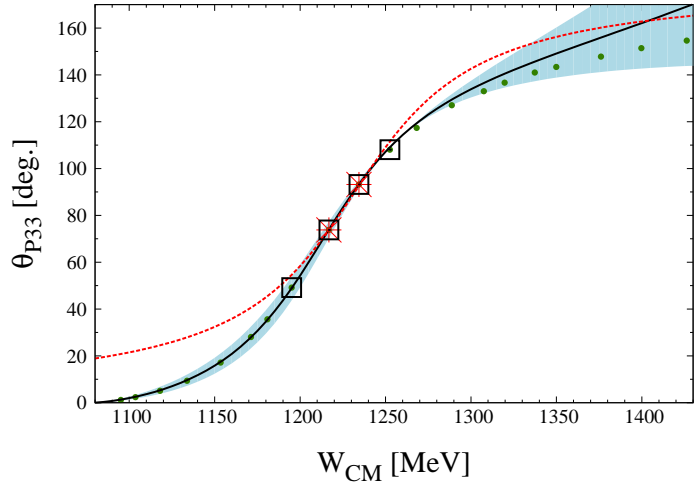}
\caption{\label{fig:p33}$P_{33}$ phase shifts (in degrees) as a function
  of $W_{\text{CM}}$ (in MeV), the CM energy including the nucleon mass.  
The EFT strict heavy-baryon expansion
at LO (NNLO)
is represented by the red dashed (black solid) line. 
The NLO curve coincides with LO.
The light-blue
band outlines the estimated systematic error of the NNLO curve.
The green dots are the
results of the GW phase-shift analysis \cite{gwpwa}. 
Points marked by a red star (black square) are inputs for LO (NNLO).
}
\end{figure}

\begin{figure}
\centering
\includegraphics[scale=0.6]{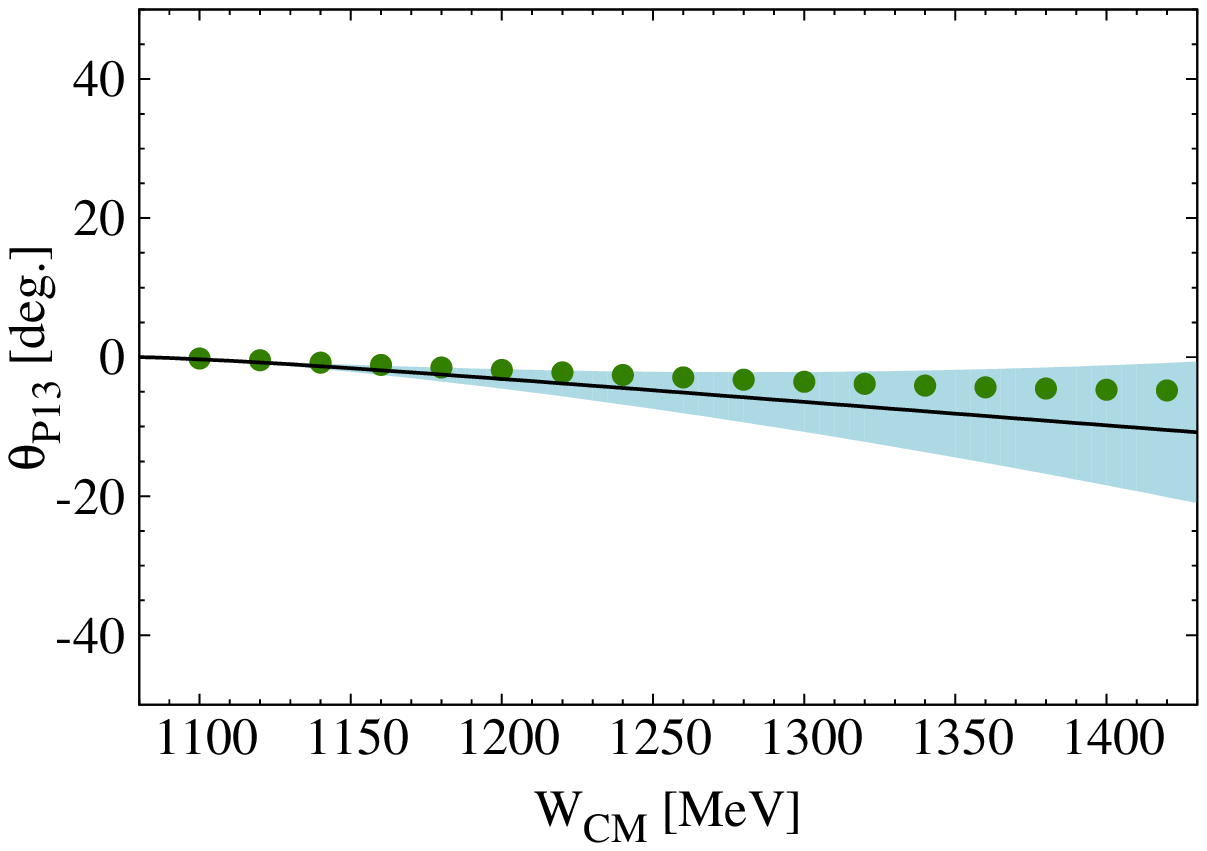}
\includegraphics[scale=0.6]{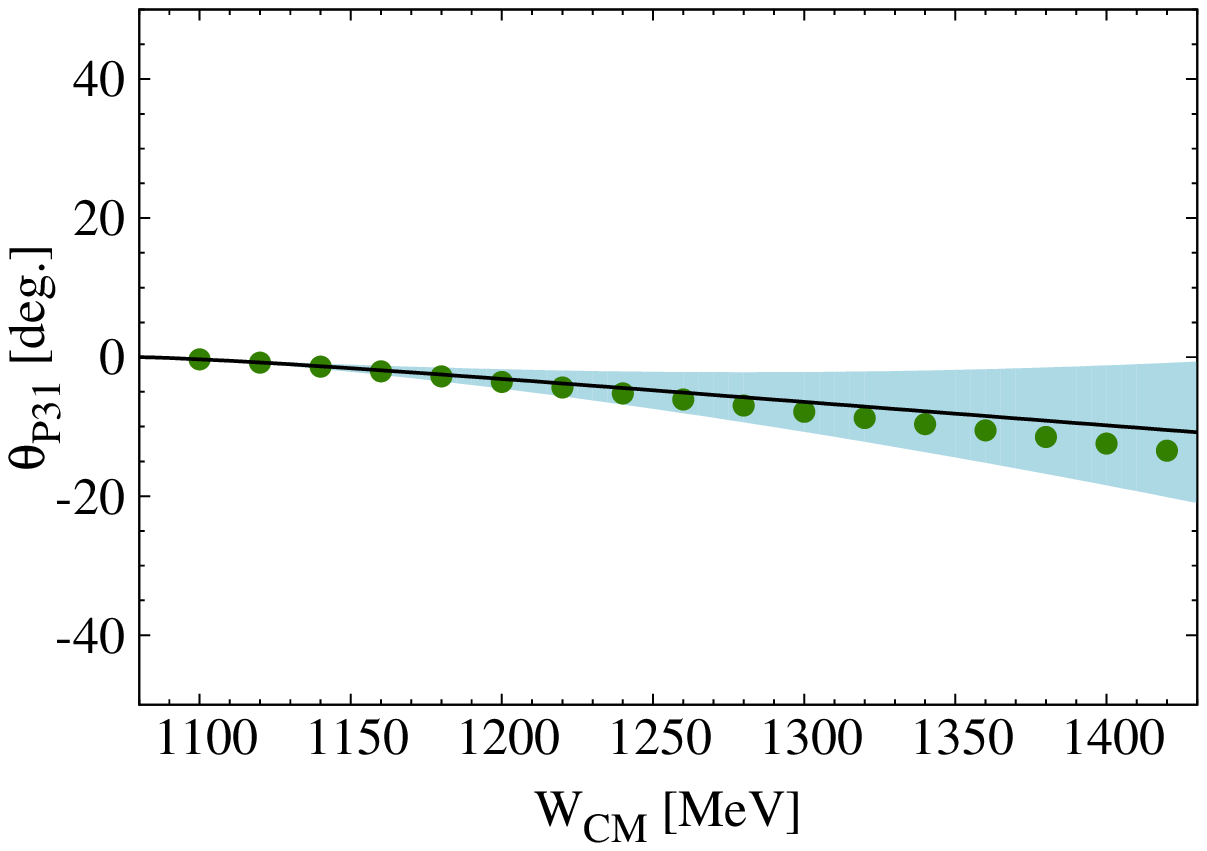}
\includegraphics[scale=0.6]{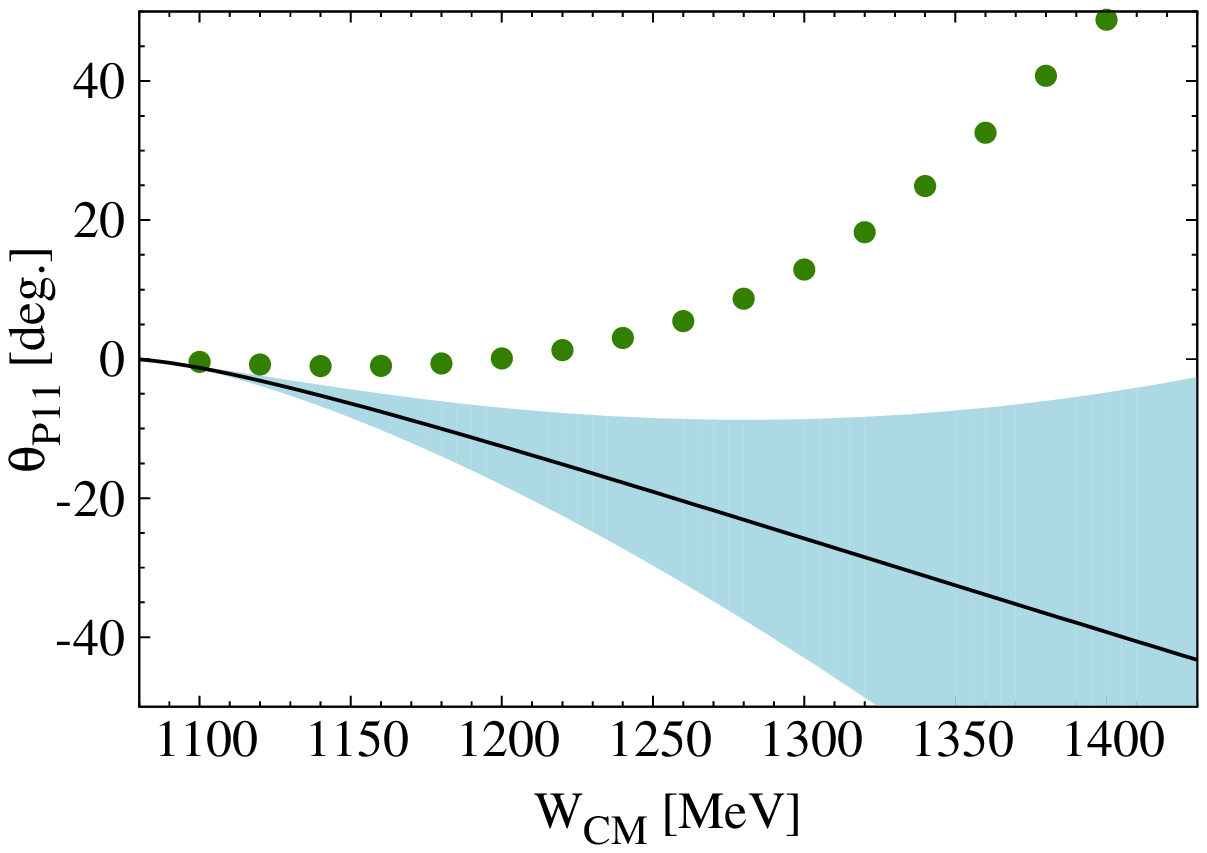}
\caption{\label{fig:otherpwaves} Predicted phase shifts  (in degrees)
in the $P_{13}$, $P_{31}$, and $P_{11}$ channels
as functions of $W_{\text{CM}}$ (in MeV),
the CM energy including the nucleon mass. 
LO and NLO vanish in these channels;
NNLO EFT results in the strict heavy-baryon expansion
are given by the black solid lines. 
The light-blue bands outline the estimated systematic errors of 
the NNLO curves.
The green dots are the
results of the GW phase-shift analysis \cite{gwpwa}. 
}
\end{figure}

The phase shifts for the strict heavy-baryon expansion 
are shown as dashed line for LO (\refeq{eqn:thtp33LO}) 
and solid line for NNLO (Eqs.~\eqref{eqn:thtp33NNLO} and \eqref{eqn:thtp131}) 
in FIGs.~\ref{fig:p33} and \ref{fig:otherpwaves}.\footnote{The LO curve
in the $P_{33}$ channel does not vanish at threshold, but, as explained 
in Sec.~\ref{sec:sewing}, the deviation from 
zero is higher-order compared to the resonance amplitude. Indeed,
the $P_{33}$ phase shift vanishes exactly at threshold starting at NNLO.}
The $P_{33}$ EFT results agree quite well with the GW PSA
past the delta resonance, and then, as expected from a momentum expansion,
start to deviate from the 
data. 
We see a reasonable convergence pattern,
and the small error band 
helps us understand why the EFT works so well in that channel: there is not
much room for the EFT to change at higher orders.
The $P_{13}$ and $P_{31}$ data are not far from the equality
predicted at NNLO, and the $h_A$ 
coming out of the $P_{33}$ fit is such that the EFT curve
provides a good description of the average phase shifts,
being slightly below the data in the $P_{13}$ channel and
above in $P_{31}$. In both cases the empirical phase shifts
are within the error bands.
In the three channels our fits are comparable to 
other unitarization methods \cite{unitarization1,unitarization2}.
Well below the resonance energy, 
our results are comparable to others at $\mathcal{O}(Q^1)$
\cite{threshold-delta1,threshold-delta2},
but not as good as $\mathcal{O}(Q^3)$ \cite{threshold-delta2},
which contains further LECs.
On the other hand,
the discrepancy with data 
in $P_{11}$ is out of the error band and thus significant. 
A similar 
discrepancy was already seen at LO in Refs.~\cite{fettes-deltaless-is,
  threshold-delta2}, where the focus was energies below the resonance.
The discrepancy could be
considered a ``small'' effect compared to the dominant $P_{33}$
amplitude. Indeed, Refs.~\cite{fettes-deltaless-is, threshold-delta2} 
found that higher-order
corrections improve the EFT description in the $P_{11}$ channel,
allowing a change in the sign of the derivative of the phase shifts at
the price of fitting more LECs.
Unitarization including the Roper \cite{unitarization1,unitarization2}
also works well throughout the energy region we consider here.

The LECs extracted from the $P_{33}$ fit are given in 
TABLE \ref{tbl:lecmix}, where they are denoted ``strict''
to emphasize that they were obtained from an
amplitude where higher orders in $Q/m_N$ were not treated
differently than higher orders in $Q/\mqcd$.
One can estimate the errors in the NNLO values 
as the variation in each LEC within which the
NNLO $P_{33}$ curve in \reffig{fig:p33} roughly stays within the
error band. This is of course not a rigorous statistical method; it
only serves to indicate how confident we are about the fitted LEC values.
This way we find $\delta$/MeV, $h_A$, and $\varkappa$ to be within
$\sim \pm 4$, $\pm 0.30$, and $\pm 0.030$, respectively, of 
the NNLO values in TABLE \ref{tbl:lecmix}.

\begin{table}
\caption{\label{tbl:lecmix}Low-energy constants
extracted at LO, NLO, and NNLO from the fits using 
the strict heavy-baryon expansion (strict)
and the partial resummation (semi).}
\begin{tabular}{|c||c c c|c c c| c|}
\hline
& \multicolumn{3}{c|}{$\delta$ (MeV)}& \multicolumn{3}{c|}{$h_A$} & $\varkappa$\\ 
& LO &NLO & NNLO & LO &NLO & NNLO & NNLO \\
\hline
\hline
strict  & 293 & 293 & 320 & 1.98 & 4.21 & 2.85 & 0.050 \\
\hline
semi    & 293 & 293 & 305 & 2.71 & 2.71 & 2.92 & 0.058 \\
\hline
\end{tabular}
\end{table}

The delta-nucleon mass splitting is related to the position
of the delta pole, which can be found by seeking the root of
\beq
S_\Delta^{-1}(E) = E - \delta + i\Gamma(E)/2 = 0\; .
\label{poleeq}
\eeq
It yields
the values given in TABLE \ref{tbl:polpos} under the label
``strict'',  
which agree fairly well with the values from
the GW PSA \cite{gwpwa} and from the Review of Particle Physics \cite{pdg}.
In addition to the value of $\delta$ in TABLE \ref{tbl:lecmix},
we find 
at LO $\Gamma(\delta)=104$ MeV and at NNLO
$\Gamma(\delta)=246$ MeV. 
The Breit-Wigner values \cite{pdg}
$\delta_\text{BW} \approx 1232 \tr{MeV} -  m_N$
and $\Gamma_\text{BW} \approx 118$ MeV 
are extracted from a fit of the form \eqref{eqn:bwback}
to data around the resonance peak.
Since it is not clear that this window coincides with
the ``Breit-Wigner window'' where our results
reduce to  \refeq{eqn:bwback}, we cannot adopt
$\delta_\text{BW}$ and $\Gamma_\text{BW}$ as 
values for $\delta$ and $\Gamma(\delta)$, even though the pole positions agree.
When other reactions are considered within our approach,
one should use  the value of $\delta$ determined above.
The large value of $\Gamma(\delta)/2$ at NNLO indicates that it is not a good
approximation for the imaginary coordinate of the pole, because
the derivatives of $\Gamma(E)$ in Eq. (\ref{poleeq}) are not small.
Since our NNLO amplitude gives a good fit of the phase shifts,
there is nothing intrinsically wrong with our $\Gamma(E)$;
what we see here is an example of the general argument given in 
Ref. \cite{gegelia} that $\Gamma(\delta)$ is scheme-dependent.

\begin{table}
\caption{\label{tbl:polpos}
Pole position of the delta resonance (in MeV)
extracted at LO and NNLO from the fits using 
the strict heavy-baryon expansion (EFT-strict)
and the partial resummation (EFT-semi)
compared with values from the GW phase-shift analysis (PSA) \cite{gwpwa}
and the Review of Particle Physics (RPP) \cite{pdg}.
EFT results at NLO are identical to LO.
}
\begin{tabular}{|c||c|c|}
\hline
& LO & NNLO \\
\hline
EFT-strict  & $1232-52i$ & $1211-50i$  \\
\hline
EFT-semi    & $1232-52i$ & $1211-48i$  \\
\hline
\hline
PSA         &   \multicolumn{2}{c|}{$1211-49.5i$} \\
\hline
RPP         &    \multicolumn{2}{c|}{$\approx 1210-50i$} \\
\hline
\end{tabular}
\end{table}

The NNLO value of $h_A$ we found is consistent with other 
sources, 
for example: 
$h_A = 1.96-2.64$ \cite{threshold-delta2},
$h_A = 2.92$ \cite{ellis-tang}, and $h_A = 2.81$ \cite{pascalutsa}.
Moreover, we obtain $h_A/g_A=2.21$, close to
the large-$N_c$ ratio $h_A/g_A= 3/\sqrt{2}$
\cite{largenc}.
The third parameter in the fit, $\varkappa$,
only appears at NNLO, and 
has several LECs embedded in itself, see \refeq{eqn:ftilde}: 
$d$ and $g^\Delta_A$, which have not yet been pinned down 
at the order we consider here.
The central value we obtain is very close to the naive estimate
$(\delta^2-m_\pi^2)/(4\pi f_\pi)^2\sim 0.05$,
but from the estimated error we see that $\varkappa$ is not
determined precisely in our fit.

Substituting the fitted values for $h_A$ and $\delta$ at NNLO in
Eqs.~\eqref{eqn:ap33} and \eqref{eqn:apnd}, we find the EFT predictions
for $P$-wave scattering volumes labeled
``strict'' in TABLE~\ref{tbl:svmix}. 
The ratio $h_A/g_A$ over-compensates for the $m_\pi/\delta$ suppression,
so that the delta contributions are not negligible. 
The EFT predictions are compared to the values extracted from
the low-energy PSA of Ref. \cite{matsinos},
which are consistent with earlier extractions.
We see good agreement for $P_{33}$ and $P_{13}$,
and $P_{31}$ is not too far off.
The $1/4$ ratio between $P_{11}$ and $P_{31}$, predicted
in \refeq{eqn:apnd}, 
is not respected well in the real world.
This is a reflection of the
discrepancy between the EFT prediction and the 
GW PSA in the $P_{11}$ channel shown in FIG. \ref{fig:otherpwaves}.

\begin{table}
\caption{\label{tbl:svmix}$P$-wave scattering volumes 
in units of $m_\pi^{-3}$ (with $m_\pi$ the charged pion mass): 
EFT in a strict heavy-baryon expansion (EFT-strict)
and
with partial resummation of relativistic corrections (EFT-semi)
compared with results from a partial-wave analysis (PSA) \cite{matsinos}.
}
\begin{tabular}{|c||c|c|c|c|}
\hline
& $a_{P_{33}}$ & $a_{P_{13}}$ & $a_{P_{31}}$ & $a_{P_{11}}$ \\
\hline
\hline
EFT-strict & $0.20$ & $-0.034$ & $-0.034$ & $-0.13$ \\
\hline
EFT-semi & $0.18$ & $-0.028$ & $-0.028$ & $-0.11$ \\
\hline
PSA & $0.2100(20)$ & $-0.03159(67)$ & $-0.04176(80)$ & $-0.0799(16)$ \\
\hline
\end{tabular}
\end{table}

As we see, the EFT in the strict heavy-baryon expansion
works pretty well for observables (except for $P_{11}$), showing in the first
two nontrivial orders good convergence to the data.
However, there are also hints that the convergence is slow
when one looks at how the parameters, particularly
$h_A$, change with order in TABLE \ref{tbl:lecmix}.
It is also a bit surprising, but not necessarily significant,
that $\Gamma(\delta)$ is found at NNLO
to be much larger than $\Gamma_{\text{BW}}$.

The strict heavy-baryon expansion is in powers of
$Q/\mqcd$ and $\delta/\mqcd$, and when $Q\sim \delta$
neither is a particularly small ratio. 
One can get a sense for the size of the corrections by
considering the case where we retain some of the 
relativistic $Q/m_N$ and $\delta/m_N$ corrections to all orders,
the approach we call semi-resummation. 
We have already discussed in Sec. \ref{sec:selfenergy}
how the slow convergence of the $\delta/m_N$ expansion
affects the width. We now turn to its effects 
in comparison to data.

\begin{figure}
\centering
\begin{tabular}{lcc}
\includegraphics[scale=0.6]{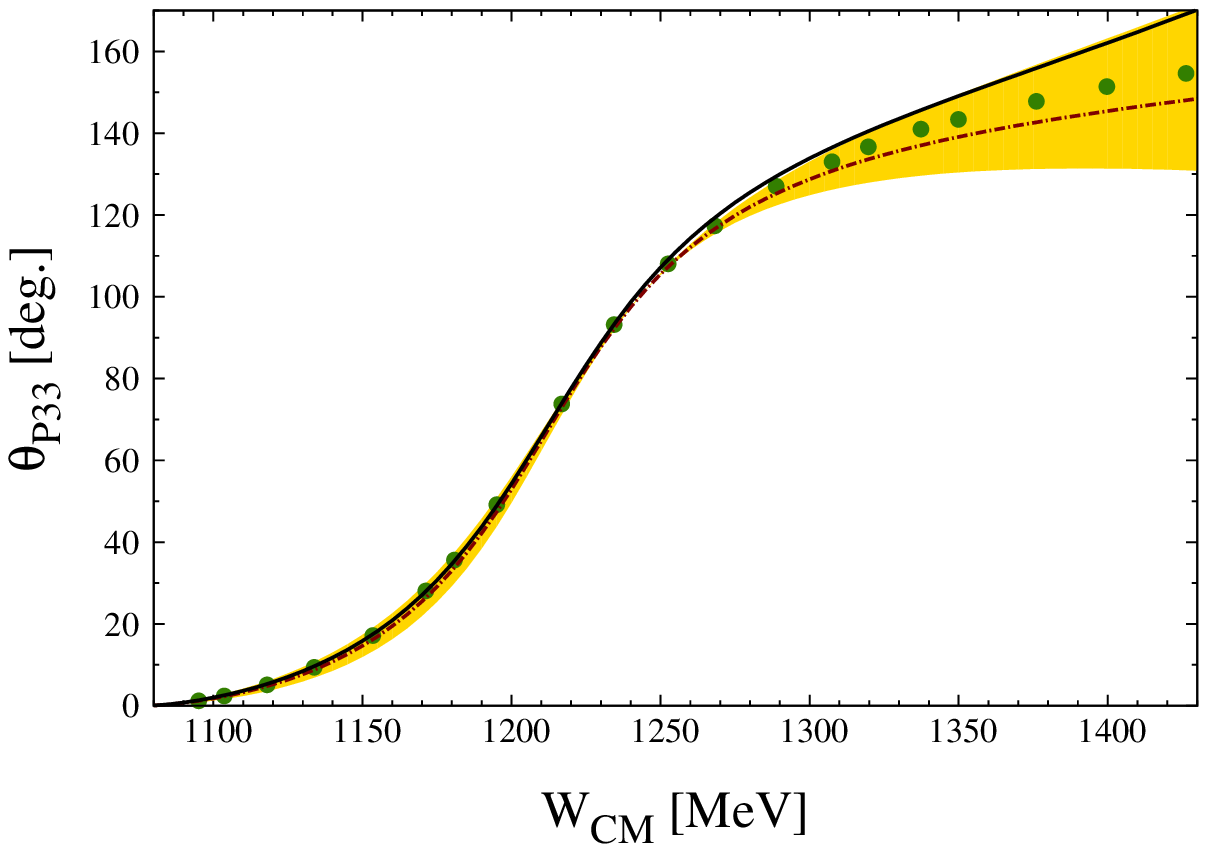}&\includegraphics[scale=0.6]{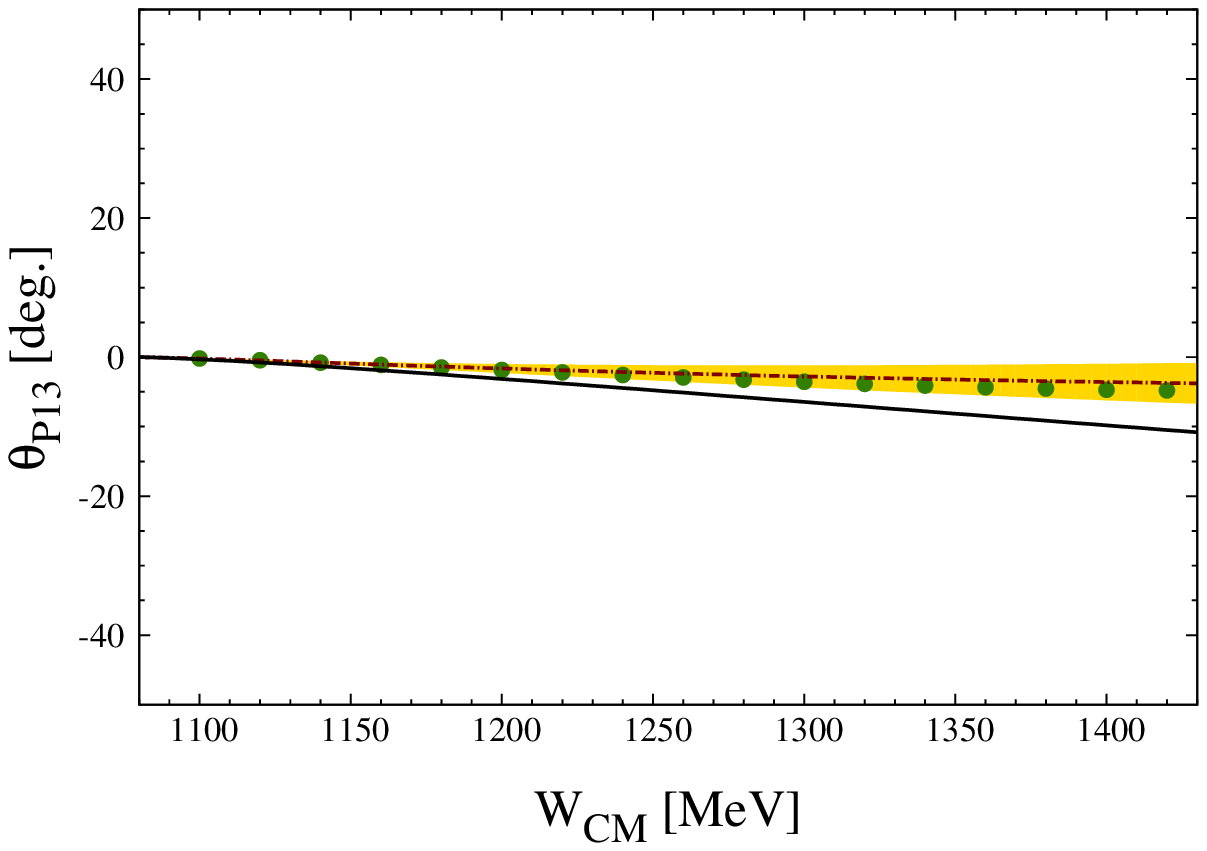}\\
\includegraphics[scale=0.6]{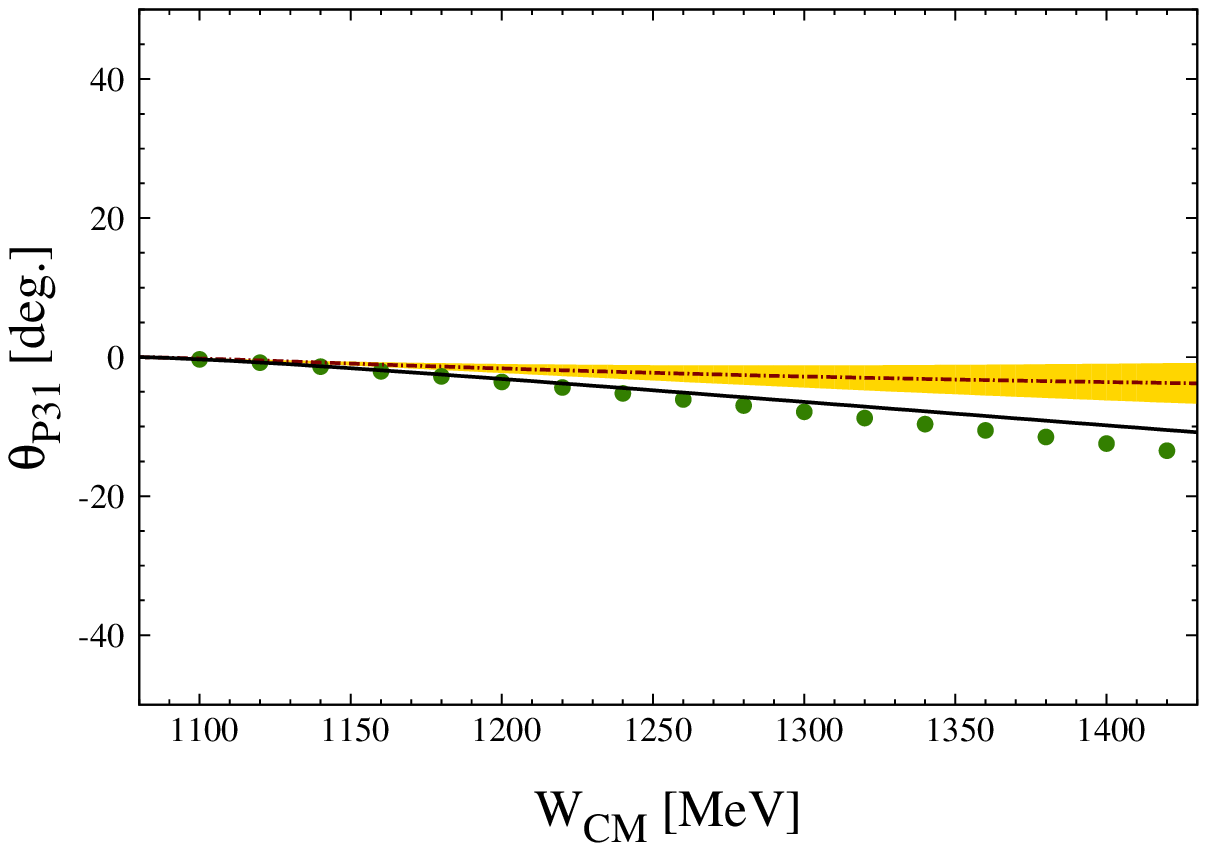}&\includegraphics[scale=0.6]{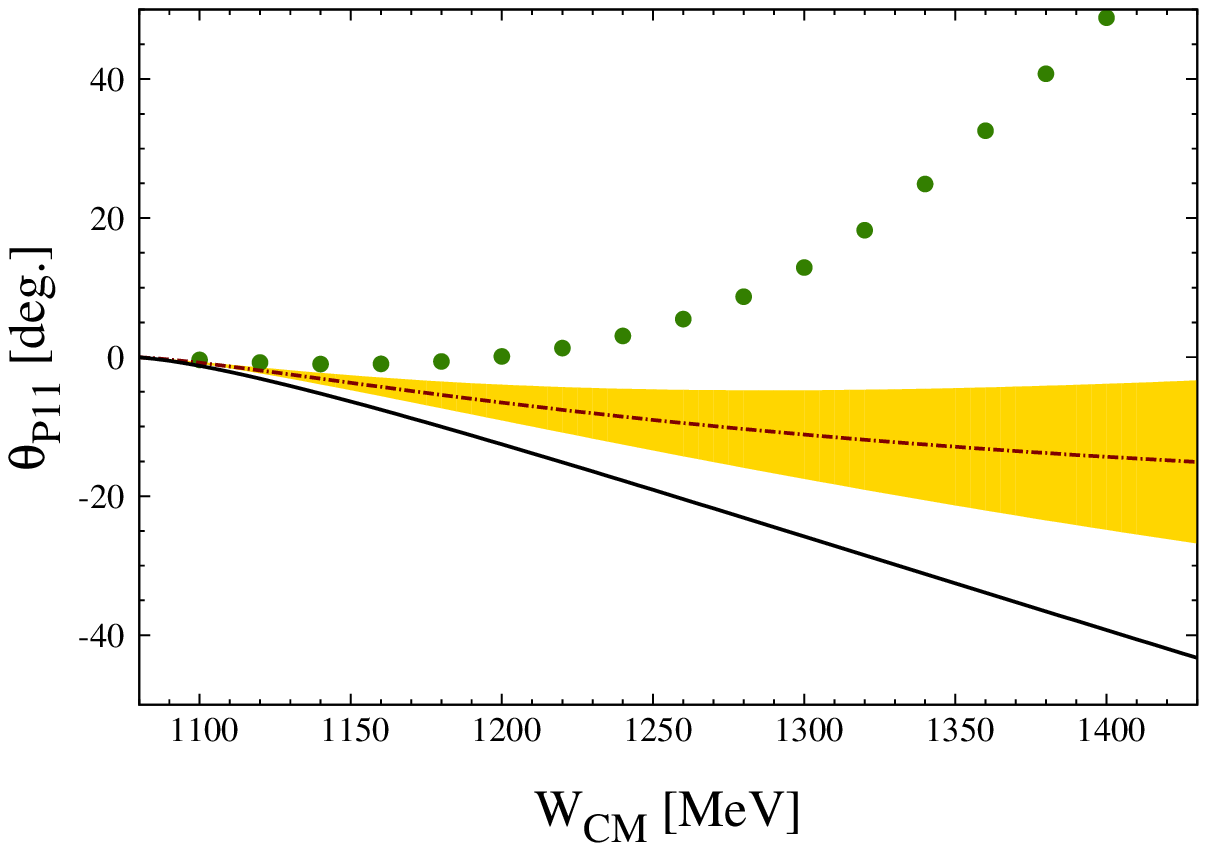}\\
\end{tabular}
\caption{\label{fig:mix}
Comparison of NNLO $P$-wave phase shifts (in degrees) as functions of 
$W_{\text{CM}}$ (in MeV), the CM energy including the nucleon mass. 
The black solid (maroon dot-dashed) lines are the
strict heavy-baryon (semi-resummed) results. 
The golden bands outline the estimated systematic errors of the 
semi-resummed expansion. 
The green dots are the
results of the GW phase-shift analysis \cite{gwpwa}. 
}
\end{figure}

At LO and NLO in the semi-resummation, there is no change in phase shifts 
compared to the strict 
heavy-baryon
expansion, and therefore also no change in the position of the delta pole;
only the relationship between width and parameters changes, 
and thus the parameters come out different.
At NNLO, the $P_{33}$ phase shifts, and with them the pole position,
change because
the dependence of the width on the energy is slightly modified.
In \reffig{fig:mix}, we 
compare the NNLO results 
(Eqs.~\eqref{eqn:thtp33NNLO} and \eqref{eqn:thtp131}) of
the partial resummation of relativistic corrections
with those of the strict heavy-baryon expansion,
exhibited before in FIGS. \ref{fig:p33} and \ref{fig:otherpwaves}.
The semi-resummation fit parameters, delta pole position,
and scattering volumes are labeled ``semi'' in TABLES
\ref{tbl:lecmix}, \ref{tbl:polpos}, and \ref{tbl:svmix}, respectively.
In this case we estimate the errors in $\delta$/MeV, $h_A$, 
and $\varkappa$ to be within $\sim \pm 3$, $\pm 0.20$, and $\pm 0.02$, 
respectively, of the values in TABLE \ref{tbl:lecmix}.
At NNLO $\Gamma(\delta)=155$ MeV.

There is a slight improvement in the $P_{33}$ phase shifts above the resonance,
and the width at resonance is closer to the Breit-Wigner value, 
without destroying the good fit at lower energies. 
The fitted parameters vary less with order than in the strict expansion,
but
at NNLO parameters in two schemes agree within $5\%$,
except for the small parameter $\varkappa$, where they differ by
$\simeq 15\%$.
The $P_{13}$ phases move in the right direction for most
energies.
In the other channels, the situation is less positive.
The $P_{31}$ phases get significantly worse.
Finally, there is improvement in the $P_{11}$ channel, but
sizable higher-order contributions are still needed to get
anywhere close to the empirical phase shifts.

There are thus regions where
each approach gets a better fit to data
than the other. 
It is important to note that the semi-resummed curves are more
or less within the error bands of their strict heavy-baryon
counterparts, as one would expect from the fact that
the differences between two methods are higher-order effects.
Overall, we do not find that either the semi-resummation or the strict
heavy-baryon works considerably better in describing data
than the other.

\section{Summary and Conclusion\label{sec:summary}}

We have shown, using $\pi N$ scattering as an example, how to
generalize standard ChPT so
that one can cope with the non-perturbative delta resonance within
EFT. Our method is similar in spirit to that 
developed in Refs. \cite{pascalutsa,morepascalutsa1,morepascalutsa2},
but differs in detail.
It can be thought of as a partial supporting argument for previous, successful
unitarization results \cite{unitarization1,unitarization2}.

ChPT has been generalized to include an explicit field for the
delta isobar. We built on earlier work  \cite{jenkins,bira-thesis}
based on the heavy-baryon
formalism, which treats the delta as a nonrelativistic
particle from the very beginning, rather than relying on
a relativistic Lagrangian of the Rarita-Schwinger field. 
We worked out
the $\mathcal{O}(Q^2/m_N^2)$ relativistic corrections to the $\pi N \Delta$
vertex within this approach. 

EFTs are based on expansions in powers of the ratio
of low to high scales, $M_{\rm lo}/M_{\rm hi}$.
The rationale to include the delta as an explicit degree of freedom
is that an expansion makes sense when 
$M_{\rm lo}\sim m_\pi \sim \delta$ and $M_{\rm hi}\sim \mqcd$.
Such an expansion is straightforward away from
regions of phase space where the delta goes on shell 
\cite{jenkins,bira-thesis,hemmertdelta}, 
but it requires change otherwise:
certain contributions are enhanced when the external 
energy is dialed to the delta-nucleon mass difference.
A way to deal with these enhancements had been explored in 
Refs. \cite{pascalutsa,morepascalutsa1,morepascalutsa2},
in which $m_\pi$ and $\delta$ were considered separate scales.
Here
we followed a similar approach to nuclear resonances \cite{resonances}
and 
constructed a power counting 
that incorporates the kinematic fine-tuning at the delta pole
while keeping $M_{\rm lo}\sim m_\pi \sim \delta$.
Our approach
resulted in a global EFT description for both the threshold and
resonance regions. 

Like other EFTs that deal with non-perturbative phenomena, 
ours captures the non-perturbative structure in LO.
Subsequently, the power counting leads to a systematic,
perturbative improvement beyond
LO.
We applied this power counting to low-energy $\pi N$
scattering, where we built the 
amplitudes up to NNLO. 
We have considered both a strict heavy-baryon expansion and 
a partial resummation of relativistic effects, which yielded 
comparable results.

Up to NNLO, we could cast our $P_{33}$ amplitude in terms of
a Breit-Wigner form with energy-dependent width and background.
Our EFT result is, however, not a mere reproduction of
\refeq{eqn:bwback}: with our approach,
{\it (1)} the size of the background can be estimated beforehand; 
{\it (2)} one need not ``cautiously'' choose the proper
energy domain for the resonance window;
and {\it (3)} there is a link between the energy dependences
in the width and background via the
parameters $\delta$, $h_A$, and $\varkappa$, 
which are then constrained by both the 
threshold and resonance data.

We fitted our $P$-wave 
amplitudes to the phase shifts given by Ref.~\cite{gwpwa}. 
With just three free
parameters, we obtained a good fit in the $P_{33}$ channel.
Contrary to an {\it ad hoc} resummation in the resonance region
\cite{ellis-tang}, we had no difficulty
preserving the unitarity condition around the resonance.
We found parameters consistent with other determinations.
With them,
the $P_{13}$ and $P_{31}$ channels come out qualitatively correct.
In the $P_{11}$ channel 
a sizable discrepancy exists between the EFT prediction, which is repulsive,
and the empirical phase shifts, which
become attractive within the region of study.

In order to improve the description of the $\pi N$ data
it is imperative to include other effects. 
One can push the calculation to next order ($\mathcal{O}(Q^2)$), 
when in channels other than
$P_{33}$ 
our framework should yield results similar to Ref. \cite{threshold-delta2}.
These lead to a better description of the phases, at least near threshold,
and in particular were found to alleviate the $P_{11}$ discrepancy.
The lack of attraction in 
this channel has been noticed long ago 
in the context of models,
and it has been attributed to the Roper resonance's fairly large width
(see, \textit{e.g.}, Ref.~\cite{ericson}). 
At $\mathcal{O}(Q^2)$
Roper effects can appear through low-energy constants.
Alternatively, Ref. \cite{threshold-delta1} made 
an attempt of incorporating an explicit Roper field in EFT but only
investigated the region that is below the delta resonance. One can use
the framework presented here to improve the description of the $P_{11}$
channel up to the energy of the Roper resonance.
 
We have thus extended ChPT in $\pi N$ scattering to the delta
region.
The EFT approach presented in this paper 
also provides the basis for a model-independent,
unified description, from threshold to past the delta
resonance without discontinuity,
of reactions involving other probes and targets,
including nuclei.
These reactions, some of which have already been successfully studied with
the pioneering power counting of Ref. \cite{pascalutsa},
present further tests of our power counting.  
A comprehensive confrontation of results from the two approaches
should indicate which one provides a more efficient organization of EFT
interactions.

\acknowledgments
We thank Daniel Phillips, Vladimir Pascalutsa, James Friar, and (at a
very early stage)
Nadia Fettes for useful discussions.
Comments on the manuscript by Daniel Phillips were greatly appreciated.
We are grateful to the following institutions for hospitality 
during the long gestation of this work:
Los Alamos National Laboratory (BwL),
the Kernfysisch Versneller Instituut at Rijksuniversiteit Groningen (UvK), 
the National Institute for Nuclear Theory at the University of Washington 
(BwL, UvK),
and the University of Arizona (BwL).
This work was supported by the US DOE (BwL, UvK)
and the Alfred P. Sloan Foundation (UvK).

\appendix*
\section{Slow-velocity boosts}
\label{app:nonrel}

A ``heavy'' particle is one that has
three-momentum $Q \ll m$, with $m$ the mass of the particle. What we
are looking for here is a Lorentz
transformation rule for a heavy-particle field that is expressed in
a perturbative fashion in powers of $Q/m \ll 1$. For $Q/m$ to remain
small, boosts of the frame under consideration
have to be in small velocities $\sim Q/m$. 
What we are 
studying here is in fact a special case of the heavy-baryon
formalism, in which the four-velocity 
label $v = (1, \vec{0})$. Since $v = (1, \vec{0})$ is sufficient
for all the processes considered in this paper, we are not
concerned with the invariance of the Lagrangian under a variation of
$v$, namely, reparameterization invariance \cite{LukMano92}.
More details will appear in Ref. \cite{bw-nonrel}.

A Poincar\'e transformation takes a spacetime point $x^{\mu}$ to
\beq
x'^{\mu} = \Lambda^{\mu}_{\nu} x^{\nu} + a^{\mu} \; ,
\eeq
with $a^{\mu}$ a four-vector representing the spacetime translation and
$\Lambda^{\mu}_{\nu}$ the Lorentz-transformation matrix. The
Poincar\'e group has ten generators: the time translation $H$, 
spatial translations $\vec{P}$, spatial
rotations $\vec{J}$, and boosts $\vec{K}$. The commutation relations
among these generators, or the Poincar\'e algebra, can be readily
found in the literature. We adopt
the notation used in Ref.~\cite{FW50}. 
The most commonly used Poincar\'e
representations are a
class of fields $\Phi_l(x)$, with $l$ a discrete index, that transform
under the Poincar\'e group as
\beq
\Phi'_{l'}(x) = M(\Lambda)_{l' l}
\Phi_l (x'') \; ,
\label{eqn:Lorrep}
\eeq
where $M(\Lambda)$ is a finite-dimension spacetime-independent
matrix that furnishes
a representation of the homogeneous Lorentz group and
\beq
x'' \equiv \Lambda^{-1} (x-a) \; .
\eeq
Therefore, for
any (non-unitary) finite-dimension representation of the homogeneous Lorentz
group, a corresponding Poincar\'e representation can be constructed using
\refeq{eqn:Lorrep}. Such representations include the well-known
Klein-Gordon scalar field, Dirac field, four-vector field, {\it etc}.

However, the Foldy-Wouthuysen (FW) representation \cite{FW50}
does not fall in the above
category. A (classical) FW field with mass $m$ and spin $s$, $\chi(t,
\vec{x})$, is an $SO(3)$ $(2s + 1)$-component spinor. The spin operators,
$\vec{s}$, are three $(2s + 1) \times (2s + 1)$ matrices, satisfying
\beq
[s_i, s_j] = i \epsilon_{i j k} s_k \; .
\eeq
The FW representation is furnished by identifying $H$, $\vec{P}$, 
$\vec{J}$, and $\vec{K}$ as follows,
\bea
H &=& \omega \; , \\
\vec{P} &=& \vec{p} \; , \\
\vec{J} &=& \vec{x} \times \vec{p} + \vec{s} \; , \\
\vec{K} &=& \frac{1}{2} \left(\vec{x} \omega + \omega \vec{x} \right)
- \frac{\vec{s} \times \vec{p}}{m + \omega} - t\vec{p} \; , \label{eqn:fwK} 
\eea
with $\vec{p} \equiv -i\vec{\nabla}$
and $\omega \equiv (m^2 +\vec{p}^{\, 2})^{\frac{1}{2}}$. 
One can check that the above definition does
satisfy the Poincar\'e algebra. In particular, an infinitesimal boost
can be written as
\beq
\chi'(t, \vec{x}) \equiv (1 - i \vec{\xi} \cdot \vec{K})\chi(t,
\vec{x}) \; ,
\eeq
where the boost is parameterized by $\vec{\xi}$, which is sometimes
referred to as rapidity and is
related to the relative velocity $\vec{V}$ between the original and
boosted frame through
\beq
\vec{\xi} = \hat{V} \textrm{tanh}^{-1} V \; .
\eeq

Because $\omega$ is a non-local operator,
the FW representation is not as useful in relativistic situations
as the Dirac field and the like. 
But the formal expansion of $\omega$ and
$\vec{K}$ in powers of $|\vec{p}|/m$ enables one to derive a
perturbative Lorentz transformation rule for the FW field,
\bea
\omega &=& m + \frac{\vec{p}^{\, 2}}{2m} - \frac{\vec{p}^{\, 4}}{8 m^3} 
+ \cdots \; ,
\label{eqn:fwomegaexp} \\
\vec{K} &=& m\vec{x} - t\vec{p} 
+ \frac{1}{4m}\left( \vec{p}^{\, 2} \vec{x} +\vec{x} \vec{p}^{\, 2} \right) 
- \frac{1}{2m} \vec{s} \times \vec{p} + \cdots 
\; .
\label{eqn:fwKexp}
\eea
We \emph{define} the LO Lorentz transformation so as to reproduce the Galilean
transformation. 
We take $|\vec{p}| \sim Q$, so that
the kinetic energy $E \equiv \omega - m$ scales as $\sim Q^2/m$.
Using the uncertainty
principle, we expect that $|\vec{x}| \sim 1/Q$ and $t \sim 1/E \sim m/Q^2$. 
Also, we power count
$\vec{s}$ as $|\vec{s}|\sim 1$.
Therefore, the LO boost generators are
\beq
\vec{K}^{(0)} = m\vec{x} - t\vec{p} \; ,
\eeq
and $\xi \sim Q/m$. 
It is worth noting that $\vec{K}^{(0)}$ does not depend on the spin
operator, 
as indeed it should not.

To remove the inert rest energy $m$, we define the heavy field
\beq
\Psi(t, \vec{x}) \equiv e^{imt}\chi(t, \vec{x}) \; ,
\eeq
for which
\beq
i \partial_t \Psi(t, \vec{x}) = (\omega - m) \Psi(t, \vec{x}) \; .
\eeq
An infinitesimal Galilean transformation of $\Psi(t, \vec{x})$ is
found to be
\beq
\Psi'(t, \vec{x}) = \left( 1 - i m \vec{\xi} \cdot \vec{x} + t
\vec{\xi} \cdot \vec{\nabla} \right) \Psi(t, \vec{x}) 
= \left\{ \left[ 1 - i m \vec{\xi} \cdot \vec{x} + 
\mathcal{O}(Q/m) \right] \Psi \right\}(t'', {\vec{x}}'') \; ,
\eeq
where
\bea
t'' &=& \left(\Lambda^{-1} x\right)_0 = t + \vec{\xi} \cdot \vec{x} \; , 
\label{eqn:tpp} \\ 
{x''}_i &=& \left(\Lambda^{-1} x\right)^i = x_i + t {\xi}_i \;. 
\label{eqn:xpp}
\eea
For simplicity, we denote the boost transformation by
\beq
\Psi \to \left[1 - i m \vec{\xi} \cdot \vec{x} \right] \Psi \; ,
\label{eqn:transPsi2}
\eeq
with the 
understanding that the left-hand side is evaluated at 
$(t,\vec{x})$ while the right-hand side at $(t'', {\vec{x}}'')$.

Going further down in the expansion in \refeq{eqn:fwKexp}, one finds the
NLO boost generators,
\beq
\vec{K}^{(1)} = \frac{1}{4m}
\left(\vec{p}^{\, 2}\vec{x} + \vec{x}\vec{p}^{\, 2}\right) 
-\frac{1}{2m}\vec{s}\times\vec{p} \; ,
\eeq
which depend on spin.
The transformation rule for the nucleon field is worked out
with $\vec{s} = \vec{\sigma}/2$,
\beq
N \to \left[1 - i m_N\vec{\xi}\cdot\vec{x} +
  \frac{i}{2m_N}\vec{\xi}\cdot\vec{\nabla} +
  \frac{1}{4m_N}\vec{\xi} \cdot
  \left(\vec{\sigma}\times\vec{\nabla}\right) + \mathcal{O}(Q^3/m_N^3) \right]
N \; . \label{eqn:bstN}
\eeq
When dealing with the delta, one needs to keep in mind that the mass-splitting
$\delta \sim Q$ in addition to the fact that the delta has spin 3/2,
\beq
\Delta \to \left[1 - i m_N\vec{\xi}\cdot\vec{x} +
  \frac{i}{2m_N}\vec{\xi}\cdot\vec{\nabla} +
  \frac{1}{2m_N}\vec{\xi} \cdot
  \left(\vec{S}^{(\frac{3}{2})}\times\vec{\nabla}\right) 
+ \mathcal{O}(Q^3/m_N^3) \right]
\Delta \; . \label{eqn:bstD}
\eeq
Although $\delta$ does not appear in the boost at this
order, it does show up in the expansion of the kinetic energy.

Our basic strategy to build an order-by-order Lorentz-invariant
Lagrangian is the following: 
{\it (i)} enumerate all the rotationally invariant
operators according to a certain scheme, \textit{e.g.}, the chiral index
$\nu$
defined in \refeq{eqn:chiind}; 
{\it (ii)} use Eqs.~(\ref{eqn:bstN}) and
(\ref{eqn:bstD}) to constrain the coefficients of these operators so
as to preserve Lorentz invariance up to the order under consideration. 
Up to index $\nu=2$ we find the terms shown in 
Eqs.~(\ref{eqn:lag0}), (\ref{eqn:lag1}), and (\ref{eqn:lag2}).
One
can explicitly check that the $\pi N \Delta$ operators,
for example, 
are Lorentz invariant in the sense that 
Lorentz-violating effects only
arise at $\nu=3$ or higher.


\begin{thebibliography}{99}

\bibitem{pdg}
C.~Amsler \textit{et al.} (Particle Data Group),
\textit{Phys. Lett.} \textbf{B667} (2008) 1.

\bibitem{colltheo}
M.L.~Goldberger and K.M.~Watson,
\textit{Collision Theory}, John Wiley \& Sons, Inc., New York (1964).

\bibitem{weinbergbook}
S.~Weinberg,
\textit{Quantum Theory of Fields}, Vol. 2, Cambridge University Press,
New York (1996).

\bibitem{gegelia}
D.~Djukanovic, J.~Gegelia, and S.~Scherer,
\textit{Phys. Rev.} \textbf{D76} (2007) 037501.

\bibitem{weinberg79} 
S.~Weinberg, 
\textit{Physica} \textbf{96A} (1979) 327;
J.~Gasser and H. Leutwyler,
\textit{Ann. Phys.} \textbf{158} (1984) 142;
\textit{Nucl. Phys.} \textbf{B250} (1985) 465.

\bibitem{BerKaiMei}
V.~Bernard, N.~Kaiser, and U.-G.~Mei{\ss}ner,
\textit{Int. J. Mod. Phys.} \textbf{E4} (1995) 193;
V.~Bernard,
\textit{Prog. Part. Nucl. Phys.} \textbf{60} (2008) 82.

\bibitem{bira99review}
U.~van Kolck,
\textit{Prog. Part. Nucl. Phys.} \textbf{43} (1999) 337;
E.~Epelbaum,
\textit{Prog. Part. Nucl. Phys.} \textbf{57} (2006) 654.

\bibitem{originalS}
S.~Weinberg,
\textit{Phys. Rev. Lett.} \textbf{17} (1966) 616;
Y.~Tomozawa,
\textit{Nuovo Cim.} \textbf{46A} (1966) 707.

\bibitem{mojzis}
M.~Moj\u{z}i\u{s},
\textit{Eur. Phys. J.} \textbf{C2} (1998) 181.

\bibitem{fettes-deltaless-is}
N.~Fettes, U.-G.~Mei{\ss}ner, and S.~Steininger,
\textit{Nucl. Phys.} \textbf{A640} (1998) 199;
N.~Fettes and U.-G.~Mei{\ss}ner,
\textit{Nucl. Phys.} \textbf{A676} (2000) 311;
\textit{Nucl. Phys.} \textbf{A693} (2001) 693.

\bibitem{otherdeltaless}
T.~Becher and H.~Leutwyler,
\textit{JHEP} \textbf{0106} (2001) 017.

\bibitem{vijay}
V.R.~Pandharipande, D.R.~Phillips, and U.~van Kolck,
\textit{Phys. Rev.} \textbf{C71} (2005) 064002.

\bibitem{jenkins}  
E.~Jenkins and A.V.~Manohar, 
\textit{Phys. Lett.} \textbf{B259} (1991) 353;
in {\it Effective Field Theories of the Standard Model},
U.-G. Mei{\ss}ner (editor), World Scientific, Singapore (1992);
E. Jenkins,
\textit{Nucl. Phys.} \textbf{B375} (1992) 561.

\bibitem{hemmertdelta} 
T.R.~Hemmert, B.R.~Holstein, and J.~Kambor,
\textit{Phys. Lett.} \textbf{B395} (1997) 89;
\textit{J. Phys.} \textbf{G24} (1998) 1831.

\bibitem{bira-thesis}
U.~van Kolck, 
Ph.D dissertation, U. of Texas (1993);
C.~Ord\'o\~nez, L.~Ray, and U.~van Kolck, 
\textit{Phys. Rev. Lett.} \textbf{72} (1994) 1982;
\textit{Phys. Rev.} \textbf{C53} (1996) 2086;
U.~van Kolck, 
\textit{Phys. Rev.} \textbf{C49} (1994) 2932;
T.D.~Cohen, J.L.~Friar, G.A.~Miller, and U.~van Kolck,
\textit{Phys. Rev.} \textbf{C53} (1996) 2661.

\bibitem{kaiserbrock}
N. Kaiser, S. Garstend\"orfer, and W. Weise,
\textit{Nucl. Phys.} \textbf{A637} (1998) 395;
H. Krebs, E. Epelbaum, and U.-G. Mei{\ss}ner,
\textit{Eur. Phys. J.} \textbf{A32} (2007) 127.

\bibitem{threshold-delta1}
A.~Datta and S.~Pakvasa,
\textit{Phys. Rev.} \textbf{D56} (1997) 4322.

\bibitem{threshold-delta2}
N.~Fettes and U.-G.~Mei{\ss}ner, 
\textit{Nucl. Phys.} \textbf{A679} (2001) 629.

\bibitem{ellis-tang}
P.J.~Ellis and H.-B.~Tang,
\textit{Phys. Rev.} \textbf{C57} (1998) 3356;
K.~Torikoshi and P.J.~Ellis,
\textit{Phys. Rev.} \textbf{C67} (2003) 015208.

\bibitem{pascalutsa} 
V.~Pascalutsa and D.R.~Phillips, 
\textit{Phys. Rev.} \textbf{C67} (2003) 055202;
V.~Pascalutsa,
\textit{Prog. Part. Nucl. Phys.} \textbf{61} (2008) 27.

\bibitem{morepascalutsa1} 
V.~Pascalutsa and M.~Vanderhaeghen, 
\textit{Phys. Rev. Lett.} \textbf{94} (2005) 102003;
\textit{Phys. Rev. Lett.} \textbf{95} (2005) 232001;
\textit{Phys. Rev.} \textbf{D73} (2006) 034003;
\textit{Phys. Rev.} \textbf{D77} (2008) 014027.

\bibitem{morepascalutsa2} 
V.~Pascalutsa, M.~Vanderhaeghen, and S.N.~Yang,
\textit{Phys. Rept.} \textbf{437} (2007) 125.

\bibitem{original}
S. Weinberg,
\textit{Phys. Lett.} \textbf{B251} (1990) 288;
\textit{Nucl. Phys.} \textbf{B363} (1991) 3.

\bibitem{unitarization1}
U.-G. Mei{\ss}ner and J.A. Oller,
\textit{Nucl. Phys.} \textbf{A673} (2000) 311.

\bibitem{unitarization2}
M.F.M. Lutz and E.E. Kolomeitsev,
\textit{Nucl. Phys.} \textbf{A700} (2002) 193.

\bibitem{resonances} 
P.F.~Bedaque, H.W.~Hammer, and U.~van Kolck,
\textit{Phys. Lett.} \textbf{B569} (2003)  159.

\bibitem{halos} 
C.A.~Bertulani, H.W.~Hammer, and U.~van Kolck, 
\textit{Nucl. Phys.} \textbf{A712} (2002)  37.

\bibitem{gwpwa}
R.A.~Arndt, W.J.~Briscoe, I.I.~Strakovsky, and R.L.~Workman,
\textit{Phys. Rev.} \textbf{C74} (2006) 045205;
R.A.~Arndt, W.J.~Briscoe, I.I.~Strakovsky, R.L.~Workman, and
M.M.~Pavan,
\textit{Phys. Rev.} \textbf{C69} (2004) 035213;
R.A.~Arndt \textit{et al.},
The SAID program, \texttt{http://gwdac.phys.gwu.edu/}.

\bibitem{matsinos}
E. Matsinos, W.S. Woolcock, G.C. Oades, G. Rasche, and A. Gashi,
\textit{Nucl. Phys.} \textbf{A778} (2006)  95.

\bibitem{Bernard:1992qa}
V.~Bernard, N.~Kaiser, J.~Kambor, and U.-G.~Mei{\ss}ner,
\textit{Nucl. Phys.}  {\bf B388} (1992) 315.

\bibitem{sakurai}
J.J.~Sakurai,
\textit{Modern Quantum Mechanics}, Addison-Wesley, USA (1994).

\bibitem{bw-nonrel}
Bingwei Long,
in preparation.

\bibitem{FW50}
L.L.~Foldy and S.A.~Wouthuysen,
\textit{Phys. Rev.} \textbf{78} (1950) 29;
L.L.~Foldy,
\textit{Phys. Rev.} \textbf{102} (1956) 568.
 
\bibitem{wnbgnonliear}
S.~Weinberg,
\textit{Phys. Rev.} \textbf{166} (1968) 1568.

\bibitem{ccwz}
S.~Coleman, J.~Wess, and B.~Zumino, 
\textit{Phys. Rev.} \textbf{177} (1969)  2239;
C.G.~Callan, S.~Coleman, J.~Wess, and  B.~Zumino, 
\textit{Phys. Rev.} \textbf{177} (1969)  2247.

\bibitem{anotherjenkins}
E. Jenkins,
\textit{Phys. Rev.} \textbf{D53} (1996) 2625.

\bibitem{beeingsavage}
S.R. Beane and M.J. Savage, 
\textit{Nucl. Phys.} {\bf A717} (2003) 104.

\bibitem{ericson}
T.~Ericson and W.~Weise,
\textit{Pions and Nuclei}, Clarendon Press, Oxford (1988).

\bibitem{Banerjee:1994bk}
M.K.~Banerjee and J.~Milana,
\textit{Phys. Rev.}  {\bf D52} (1995) 6451;
C. Hacker, N. Wies, J. Gegelia, and S. Scherer,
\textit{Phys. Rev.} {\bf C72} (2005) 055203;
A.~Semke and M.F.M.~Lutz,
\textit{Nucl. Phys.}  {\bf A778} (2006) 153.

\bibitem{largenc}
A. Manohar, 
hep-ph/9802419.

\bibitem{BoM}
U. van Kolck, M.C.M. Rentmeester, J.L. Friar, T. Goldman, and J.J. de Swart,
{\it Phys. Rev. Lett.} {\bf 80} (1998) 4386;
J.J. de Swart, M.C.M. Rentmeester, and R.G.E. Timmermans,
{\it PiN Newslett.} {\bf 13} (1997) 96;
U. van Kolck, J.L. Friar, and T. Goldman,
{\it Phys. Lett.} {\bf B371} (1996) 169.

\bibitem{LukMano92}
M.E.~Luke and A.V.~Manohar,
{\it Phys. Lett.} {\bf B286} (1992) 348.


\end{thebibliography}
\end{document}